\documentclass[reqno]{amsart}

\usepackage{amsthm,amsmath,amsfonts,amssymb,epsfig,graphicx,latexsym,float,color}

\newtheorem{definition}{Definition}

\newtheorem{remark}[definition]{Remark}

\renewcommand{\theequation}{\arabic{section}.\arabic{equation}}

\renewcommand{\thefigure}{\arabic{section}.\arabic{figure}}

\begin{document}

\title{Whipping of Electrified Visco-Capillary Jets in Airflows}

\author[Arne et al.]{Walter Arne$^{1}$}
\author[]{Nicole Marheineke$^{2}$}
\author[]{Miguel P\'erez-Saborid$^{3}$}
\author[]{Javier Rivero-Rodr\'iguez$^{4}$}
\author[]{Raimund Wegener$^{1}$} 
\author[]{Manuel Wieland$^{1}$}

\date{\today\\
$^1$ Fraunhofer ITWM, Fraunhofer Platz 1, D-67663 Kaiserslautern, Germany\\
$^2$ FAU Erlangen-N\"urnberg, Lehrstuhl Angewandte Mathematik I, Cauerstr.~11, D-91058 Erlangen, Germany\\
$^3$ Area de Mecanica de Fluidos, Departamento de Ingeneria Aeroespacial y Mecanica de Fluidos, Universidad de Sevilla, Avenida de los Descubrimientos s/n, 41092, Sevilla, Spain\\
$^4$ TIPs Laboratory, Universit\'{e} Libre de Bruxelles C.P. 165/67, 50 av. F. Roosevelt, 1050 Bruxelles, Belgium
}

\begin{abstract}
An electrified visco-capillary jet shows different dynamic behavior, such as cone forming, breakage into droplets, whipping  and coiling, depending on the considered parameter regime. The whipping instability that is of fundamental importance for electrospinning has been approached by means of stability analysis in previous papers. In this work we alternatively propose a model framework in which the instability can be computed straightforwardly as the stable stationary solution of an asymptotic Cosserat rod description. For this purpose, we adopt a procedure by Ribe (Proc.\ Roy.\ Soc.\ Lond.\ A, 2004) describing the jet dynamics with respect to a frame rotating with the a priori unknown whipping frequency that itself becomes part of the solution. The rod model allows for stretching, bending and torsion, taking into account inertia, viscosity, surface tension, electric field and air drag. For the resulting parametric boundary value problem of ordinary differential equations we present a continuation-collocation method. On top of an implicit Runge-Kutta scheme of fifth order, our developed continuation procedure makes the efficient and robust simulation and navigation through a high-dimensional parameter space possible. Despite the simplicity of the employed electric force model the numerical results are convincing, the whipping effect is qualitatively well characterized.
\end{abstract}

\maketitle

\noindent
{\sc Keywords.} electrified jets, lateral instabilities, electrospinning, viscous Cosserat rod model, parametric boundary value problem, homotopy method\\
{\sc AMS-Classification.} 34B08, 65Lxx, 76-XX

\section{Introduction}

\begin{figure}
\includegraphics[height=5.5cm]{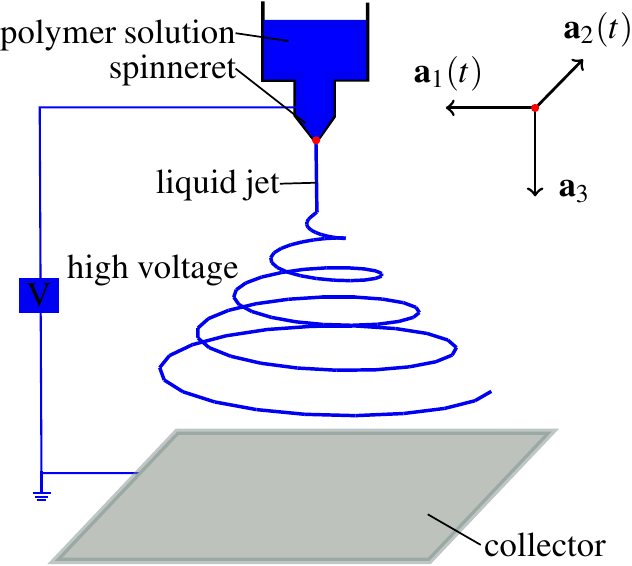}\hfill
\includegraphics[height=5.5cm]{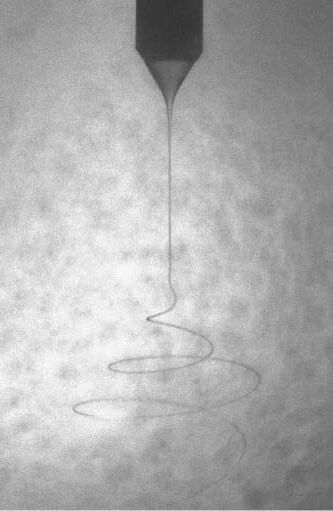}\hfill
\includegraphics[height=5.5cm]{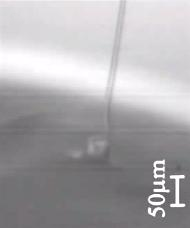}
\caption{\label{fig:photos}\emph{From left to right:} Sketch of an electrospinning or electrospray device; whipping instability in an electrified jet of glycerine in a bath of hexane (courtesy of A.\ Gomez-Marin); coiling of an electrified liquid jet (courtesy of G.\ Riboux)}
\end{figure}

The interaction of an intense electrical field with the interface between a conducting liquid and a dielectric medium has been known to exist since Gilbert \cite{gilbert:b:1600} reported in 1600 the formation of a conical meniscus when an electrified piece of ambar was brought close enough to a water drop. The deformation of the interface is caused by the force that the electric field exerts on the net surface charge induced by the field itself. This phenomenon is at the base of modern devices for the production of micro- and nanostructures of interest in several technological fields \cite{loscertales:p:2002,li:p:2004a}.  As schematized in Fig.~\ref{fig:photos} (left), these devices consist essentially of a high-voltage power supply, a metallic needle (spinneret) and a grounded collector (counter-electrode). The metallic needle is connected to a syringe pump through which a conducting liquid can be fed at a constant and controllable rate. When a high voltage is applied, the electric field induces an electric current in the liquid that accumulates electric charge at the surface and causes an electric force that elongates the pendant drop at the needle's exit in the direction of the field. It is observed that if the field strength is below a certain threshold value the balance of electrostatic and surface tension forces gives rise to a motionless conical shape commonly known as the Taylor cone \cite{taylor:p:1964}. However, above the threshold, the large electrostatic forces concentrated near the cone tip overcome the surface tension stresses and force the ejection of an electrified liquid jet from the cone tip. For certain values of the applied voltage and imposed liquid flow rate, the jet emanating from the cone tip is stationary and breaks into spherical droplets at some distance downstream due to axisymmetric Rayleigh-Plateau (varicose) instabilities corrected to account for the presence of surface charge. This so-called cone-jet mode forms the basis of the electrospray technique \cite{fenn:p:1989,mora:p:1994} for generating small monodisperse drops with great applications in fine coatings, synthesis of powders, micro- and nanocapsules, etc. However, non-symmetric perturbation modes can also grow due to the net charge carried by the jet. Indeed, if a small portion of the charged jet moves slightly off axis, the charge distributed along the rest of the jet will push that portion farther away from the axis according to Earnshaw's theorem, thus leading to a lateral instability known as whipping or bending instability. If the growth rate associated to this whipping instability is larger than that associated to varicose jet break-up -- as may happen, for example, for sufficiently high values of the applied voltage or of the liquid viscosity --,  the off-axis movement of the jet becomes the most significant aspect of its evolution, see Fig.~\ref{fig:photos} (middle). The whipping mode manifests itself in the form of fast and violent slashes which give rise to very large tensile stresses and to a dramatic jet thinning. This is of fundamental importance in the electrospinning process \cite{doshi:p:1995,hohman:p:2001,li:p:2004,reneker:p:2000,shin:p:2001,yarin:p:2001}, where micro- or nanofibers of a polymeric fluid are produced by solidification of the jet issuing from the Taylor cone before it breaks up into droplets. The reduction of the jet diameter, that is typically several orders of magnitude makes the electrospinning technique very competitive with other existing ones (such as phase separation or self-assembly) and subject of research.  Another mode, the so-called coiling mode \cite{kim:p:2010}, is observed when the ground electrode is located sufficiently close to the needle's exit so that the liquid jet reaches the plate before being set into chaotic motion by the whipping instability, see Fig.~\ref{fig:photos} (right). The situation is then the same as if a thin stream viscous fluid such as honey is poured onto a surface from a certain height \cite{ribe:p:2004, ribe:p:2017, ribe:p:2006}. Rather than approaching the surface vertically the jet builds on it a helical structure which resembles a pile of coiled rope. 

In the last decades intensive work has been spent on experimental studies of electrospinning, whose complexity makes empirical determination of the effects of the parameters very difficult \cite{shin:p:2001,li:p:2004}.  Theoretical models were developed to predict the behavior of the charged jet, see the trend-setting papers \cite{reneker:p:2000, hohman:p:2001a, feng:p:2002}. The models were set up in terms of balance equations on basis of slender-body descriptions for the jet, including the electric field and surface tension, differing in the constitutive laws, such as linear and nonlinear, Newtonian and upper-convected Maxwell, models for  viscous and viscoelastic jet behavior, etc. Also effects like evaporation and solidification have been taken into account \cite{yarin:p:2001}. It turned out \cite{thompson:p:2007} that the final jet radius is mainly determined by the volumetric charge density, the distance from the nozzle to the collector, the orifice radius, the relaxation time and the viscosity. The external field and mutual electric interaction of multiple charged jets influence the jet path and evolution \cite{theron:p:2005}. The whipping instability, which causes the jet thinning, was documented and analyzed in \cite{reneker:p:2000, yarin:p:2001} by applying perturbation frequencies. On top of a linear stability analysis operating diagrams for different polymer solvents were developed in \cite{hohman:p:2001a,hohman:p:2001}. The instability was also explored in temporal stability analysis in \cite{li:p:2008,li:p:2009}. Numerical simulations are given in \cite{zeng:p:2006}. However, it must be realized that the nature of the whipping regime makes it very difficult to unravel its detailed structure. Recent experiments show that there are some circumstances which greatly enhance the parametric range for which the bending leads to a stable steady-state helicoidal structure with a constant opening angle as, for example, when the conducting liquid is surrounded by a dielectric bath \cite{riboux:p:2011} or by another coflowing liquid \cite{guerrero:p:2014} (see Fig.~\ref{fig:photos}, middle). The liquid bath makes the frequency of the jet oscillations several orders of magnitude lower than those found in typical electrospinning experiments in air \cite{hohman:p:2001,reneker:p:2008}. In the phenomenological study on a glycerine jet in a hexane bath \cite{riboux:p:2011} the whipping instability (frequency, amplitude, wavelength) is characterized in terms of flow rate, applied voltage and electric conductivity.

The observation of the stable steady-state helical structure motivates the idea of this paper. We propose a model framework in which the whipping instability can be computed straightforwardly as the stable stationary solution of an asymptotic Cosserat rod description. For this purpose we formulate the jet's whipping as a stationary process by introducing a frame that rotates with the a priori unknown whipping frequency, so the frequency becomes part of the solution. A similar transformation has been used in the investigation of viscous rope coiling \cite{ribe:p:2004}. In the whipping instability this results in a parametric boundary value problem of ordinary differential equations which describes the jet behavior in terms of 19 variables in dependence on viscosity, surface tension, external electric field, self-repulsion of the induced charges and the resistance of the surrounding fluid. The used model for the electric and capillary forces is taken from \cite{yarin:p:2001}, the material and geometrical models for the incompressible viscous rod come from the work \cite{ribe:p:2004} on viscous rope coiling and \cite{arne:p:2011, arne:p:2010} on rotational spinning. For the resistance of the jet surrounding airflow we employ a drag model developed in \cite{marheineke:p:2011}. For the numerical treatment of the problems we propose a continuation algorithm that allows the efficient and robust simulation and automatic navigation through a high-dimensional parameter space. The underlying collocation is performed with a Lobatto IIIa formula (implicit Runge-Kutta scheme of fifth order), and the resulting nonlinear system is solved with a Newton method. We explore our framework in a parameter study. Although the model is physically comparatively simple in the electric force model, the numerical results are very convincing. They show qualitatively well the characteristic jet behavior in terms of whipping frequency, elongation and throwing range. The presented model framework is able to give a quantitative explanation for the observed strong jet thinning. Indeed, the analysis of the periodic equilibrium states in combination with a temporal global stability analysis reveal the helical structure in the whipping instability in more details as it is done in previous works. In addition, it enables the derivation of an analytical solution for the 'longtime' behavior of the jet.

The paper is structured as follows. Starting with a short introduction into the viscous Cosserat theory, we derive the stationary model framework for the jet's whipping in Sec.~\ref{sec:2}. In Sec.~\ref{sec:3} we present a continuation-collocation method for the numerical solution of the parametric boundary value problem and demonstrate the computational efficiency of the proposed algorithm. Numerical results are shown and discussed in Sec.~\ref{sec:4} with respect to parameter studies and a detailed investigation of the whipping effect. In particular, an analytical solution for the jet's 'longtime' behavior is stated. Moreover, we include a temporal stability analysis of the underlying transient Cosserat rod model in App.~\ref{appendixA}. Appendix~\ref{appendixB} gives details to apparent boundary layers in the numerical solutions.

\setcounter{equation}{0} \setcounter{figure}{0}
\section{Viscous Cosserat Rod Model}\label{sec:2}

\subsection{Electrified visco-capillary jet in airflow} 
As a jet is a slender long object, its dynamics can be reduced to an one-dimensional description by averaging the underlying balance laws over its cross-sections. In the special Cosserat rod theory there are two constitutive elements: a curve $\mathbf{r}:\mathcal{D}\rightarrow\mathbb{E}^3$  specifying the jet position (e.g.\ midline) and an orthonormal director triad $\{\mathbf{d_1},\mathbf{d_2},\mathbf{d_3}\}:\mathcal{D}\rightarrow\mathbb{E}^3$ that is attached to the curve and characterizes the orientation of the cross-sections in the three-dimensional Euclidean space $\mathbb{E}^3$. The Euclidian space $\mathbb{E}^3$ can be identified with $\mathbb{R}^3$ by choosing a basis. We particularly use the domain $\mathcal{D}=\{(s,t)\in (\mathbb{R}_0^+)^2 \}$ where $t$ is the time and $s$ is the arc length parameter to impose an Eulerian (spatial) description. In the following we consider an electrified visco-capillary jet with circular-shaped cross-sections surrounded by an airflow. To describe its dynamics we proceed from the incompressible viscous Cosserat rod model of \cite{arne:p:2011, arne:p:2010} that was derived for jets in rotational spinning processes on the basis of the work \cite{ribe:p:2004} on viscous rope coiling. The rod model consists of balances for mass (cross-section), linear and angular momentum and allows for stretching, bending and torsion. The electric and capillary forces are included according to \cite{yarin:p:2001}, whereas the air drag model is taken from \cite{marheineke:p:2011}.
The resulting rod system is given by the following four kinematic and three dynamic equations
\begin{equation}\label{eq:rod}
\begin{aligned}
\partial_s \mathbf{r} = \mathbf{d_3},  \qquad
\partial_s \mathbf{d_i} = \boldsymbol{\kappa} \times \mathbf{d_i}, &\qquad
\partial_t \mathbf{r} = \mathbf{v} - u \mathbf{d_3}, \qquad
\partial_t \mathbf{d_i} = (\boldsymbol{\omega}-u\boldsymbol{\kappa}) \times \mathbf{d_i},\\
\partial_t A + \partial_s (u A) &= 0, \\
\rho \partial_t (A \mathbf{v}) + \rho \partial_s (u A\mathbf{v}) &= \partial_s \mathbf{n} + \mathbf{f}_{ca} + \mathbf{f}_{el} +  \mathbf{f}_{air},\\
\rho \partial_t (\mathbf{J} \cdot \boldsymbol{\omega}) + \rho \partial_s (u \mathbf{J} \cdot \boldsymbol{\omega}) &= \partial_s \mathbf{m} + \mathbf{d_3}\times \mathbf{n}
\end{aligned}
\end{equation}
that are supplemented with a relation between cross-sections and moments of inertia, with viscous material laws and with outer forces
\begin{align*}
\quad A={\pi} a^2, & \qquad J =\frac{\pi}{4} a^4, \qquad \mathbf{J}=J\mathbf{P}_2,\\ 
\mathbf{n} \cdot \mathbf{d_3} = 3 \mu A \partial_s u , &\qquad \mathbf{m} = 3 \mu J \mathbf{P}_{2/3} \cdot \partial_s \boldsymbol{\omega},\\
\mathbf{f}_{ca} = \pi \gamma \partial_s (a \mathbf{d_3}),& \qquad 
\mathbf{f}_{el} = 2 \pi a\sigma \left(\mathbf{E}- \frac{a \sigma}{2 \varepsilon_{p}} \log \left(\frac{H}{a}\right) \boldsymbol{\kappa} \times \mathbf{d_3}\right), \qquad \mathbf{f}_{air} = \frac{\mu_\star^2}{2a\rho_\star}\mathbf{F}\bigg(\mathbf{d_3},-\frac{2a\rho_\star}{\mu_\star}\mathbf{v}\bigg)
\end{align*}
with the scaling tensors $\mathbf{P}_{k}=\mathbf{d_1} \otimes \mathbf{d_1}  + \mathbf{d_2} \otimes \mathbf{d_2} + k \, \mathbf{d_3}  \otimes \mathbf{d_3} $, $k\in \mathbb{R}$. 

The unknowns of the system (\ref{eq:rod}) are the jet curve $\mathbf{r}$, triad $\{\mathbf{d_1},\mathbf{d_2},\mathbf{d_3}\}$, curvature $\boldsymbol{\kappa}$, cross-section $A$, linear $\mathbf{v}$ and angular $\boldsymbol{\omega}$ velocities as well as the convective speed $u$ and the normal contact force components $\mathbf{n}\cdot \mathbf{d_i}$, $i=1,2$. The curve and the triad are coupled by $\partial_s \mathbf{r}=\mathbf{d_3}$.  In the chosen Eulerian description this relation for the jet tangent contains the arc length parameterization and the generalized Kirchhoff constraint that allows for stretching and prevents shearing as jet deformation. In this context the convective speed (parameter speed) $u$ can be viewed as the Lagrange multiplier (unknown) to the arc length parametrization. The further kinematic equations relate the jet curve and the triad to the curvature $\boldsymbol{\kappa}$ and the linear and angular velocities $\mathbf{v}$, $\boldsymbol{\omega}$. The mass density $\rho$ is considered to be constant and the cross-section $A$ to be circular-shaped of radius $a$. The coupling of the angular momentum line density with the moment of inertia $\mathbf{J}$ preserves the jet's incompressibility. The tangential contact force $\mathbf{n}\cdot \mathbf{d_3}$ and the couple $\mathbf{m}$ are specified by a linear material law in the strain rate variables with dynamic viscosity $\mu$, whereas the normal force components are the Lagrange multipliers to the generalized Kirchhoff constraint. The acting outer line force densities consist of the capillary line force density $\mathbf{f}_{ca}$ with surface tension coefficient $\gamma$, the electric line force density $\mathbf{f}_{el}$ and the aerodynamic line force density $\mathbf{f}_{air}$. The last is based here on a stationary non-moving flow situation where the density $\rho_\star$ and the dynamic viscosity $\mu_\star$ of the air are considered to be constant.

The electric forces are assumed to be split into two parts, the effects due to the external unperturbed electric field $\mathbf{E}$ and the ones due to the Coulomb interactions of the induced charges on the jet. The self-repulsion is modeled by help of a local interaction approximation \cite{yarin:p:2001} in terms of the surface charge density $\sigma$, the permittivity $\varepsilon_{p}$ and the distance between nozzle and counter-electrode  (device height) $H$ as typical (cut off) length. The external electric field depends on the applied voltage $\Phi$ and the device geometry. We assume it to be constant, i.e., $\mathbf{E}=E\mathbf{a_3}$ with $E= \Phi /H $ and $\|\mathbf{a_3}\|=1$, see Fig.~\ref{fig:photos} (left) for the device-specific direction $\mathbf{a_3}$. Nevertheless, the Cosserat rod model \eqref{eq:rod} is still not closed, because the surface charge density $\sigma$ is in no situation a constant parameter, but an unknown in the problem. Instead of $\sigma$, we consider the electric current $I$ resulting from convection and conduction, i.e., $I=2\pi a \sigma u +\pi a^2 \lambda \mathbf{E} \cdot \mathbf{d_3}$ with the jet's conductivity $\lambda$. In the intended transition to stationarity the current $I$ as well as the flow rate $Q=Au$ become constant. Whereas the flow rate is prescribed at the nozzle, the current is still unknown but can be measured in experiments. In literature different phenomenological relations of the form $I\sim Q^m \Phi^n$, $m,n\in \mathbb{R}$ are documented, see e.g.\ \cite{mora:p:1994, hohman:p:2001, shin:p:2001, theron:p:2004, riboux:p:2011}. The powers vary with the material properties and device geometry, the proportionality constant crucially depends on the experimental conditions.

The dimensionless air drag function 
\begin{align*}
\mathbf{F}(\boldsymbol{\tau},\mathbf{w}) = w_\nu r_\nu(w_\nu)\boldsymbol{\nu} + w_\tau r_\tau(w_\nu)\boldsymbol{\tau}
\end{align*}
is expressed in terms of the tangential $w_\tau = \mathbf{w}\cdot\boldsymbol{\tau}$ and normal velocity components $w_\nu=\sqrt{\mathbf{w}\cdot\mathbf{w}-w_\tau^2}$ with normal vector $\boldsymbol{\nu} = (\mathbf{w}-w_\tau\boldsymbol{\tau})/w_\nu$. We particularly use the regularized air resistance coefficients $r_\nu$, $r_\tau$ given in \cite{marheineke:p:2011}
\begin{align*}
r_\nu(w_\nu) &= \begin{cases}
\sum_{j=0}^3 q_{\nu,j}w_\nu^j,\quad & w_\nu < w_0,\\
\frac{4\pi}{S(w_\nu)}\big(1-\frac{S^2(w_\nu)-S(w_\nu)/2+5/16}{32S(w_\nu)}w_\nu^2\big),\qquad &w_0 \leq w_\nu < w_1,\\
w_\nu\exp\big(\sum_{j=0}^3 p_{\nu,j}\log^j(w_\nu)\big), &w_1 \leq w_\nu \leq w_2,\\
2\sqrt{w_\nu}+0.5w_\nu, &w_2 < w_\nu,
\end{cases}\\
r_\tau(w_\nu) &= \begin{cases}
\sum_{j=0}^3 q_{\tau,j}w_\nu^j,\quad &w_\nu < w_0,\\
\frac{4\pi}{(2S(w_\nu)-1)}\big(1-\frac{2S^2(w_\nu)-2S(w_\nu)+1}{16(2S(w_\nu)-1)}w_\nu^2\big),\qquad &w_0 \leq w_\nu < w_1,\\
w_\nu\exp\big(\sum_{j=0}^3 p_{\tau,j}\log^j(w_\nu)\big), &w_1 \leq w_\nu \leq w_2,\\
2\sqrt{w_\nu}, &w_2 < w_\nu
\end{cases}
\end{align*}
with transition points $w_0 =2\exp\big(2.0022-{4\pi}/{r_\nu^S}\big)$, $w_1 = 0.1$, $w_2 = 100$, the Stokes limits
\begin{align*}
r_\nu^S = \frac{4\pi}{\log(4/\delta)}-\frac{\pi}{\log^2(4/\delta)},\qquad r_\tau^S = \frac{2\pi}{\log(4/\delta)}+\frac{\pi/2}{\log^2(4/\delta)}
\end{align*}
the function $S(w_\nu) = 2.0022 - \log(w_\nu)$ and the regularization parameter $\delta = 3.5\cdot 10^{-2}$. The other parameters $p_{k,j}$ and $q_{k,j}$ ($k \in \{\nu,\tau\}$, $j\in\{0,1,2,3\}$) ensure smoothness and are
\begin{alignat*}{6}
p_{\nu,0} &= 1.6911, \qquad &p_{\nu,1} &= -6.7222\cdot 10^{-1},\qquad
&p_{\nu,2} &= 3.3287\cdot 10^{-2},\qquad &p_{\nu,3}&=3.5015\cdot 10^{-3},\\
p_{\tau,0} &= 1.1552,\qquad &p_{\tau,1} &= -6.8479\cdot 10^{-1},\qquad
&p_{\tau,2} &= 1.4884\cdot 10^{-2},\qquad &p_{\tau,3}&=7.4966\cdot 10^{-4},
\end{alignat*}
\begin{align*}
q_{k,0} &= r_k^S,\quad q_{k,1} = 0,\quad q_{k,2} = \frac{3r_k(w_0)-w_0r_k'(w_0)-3r_k^S}{w_0^2},\quad q_{k,3} &= \frac{-2r_k(w_0)+w_0r_k'(w_0)+2r_k^S}{w_0^3}.
\end{align*}

\subsection{Jet's whipping} 
In the whipping regime the electrified visco-capillary jet forms a helical structure, see for example the experiments in \cite{hohman:p:2001, yarin:p:2001, riboux:p:2011} and Fig.~\ref{fig:photos} (middle). In previous works stability analysis has been performed using classical perturbation theory \cite{hohman:p:2001,reneker:p:2000, yarin:p:2001, li:p:2009}. We also find a strong connection between the whipping instability observed in experiments and the unstable solutions of our electrospinning model (\ref{eq:rod}), for details we refer to the temporal stability analysis in App.~\ref{appendixA}. However, our main idea in this paper is different. To explore the instability numerically, we formulate the jet's whipping as the stationary solution of the Cosserat rod model \eqref{eq:rod}. For this purpose we consider a spun jet of certain -- a priori unknown -- length $L$ with stress-free end.  At the nozzle it is straight. Because of the electric field $\mathbf{E}=E\mathbf{a_3}$, we have a fixed predominant direction in the device, i.e., $\mathbf{a_3}=\mathbf{d_3}(0,t)$ is the jet tangent at the nozzle $s=0$ for all times $t$.  We introduce a time-dependent outer basis $\{\mathbf{a_1}(t),\mathbf{a_2}(t),\mathbf{a_3}\}$, $\partial_t \mathbf{a_i}=\boldsymbol{\Omega}\times \mathbf{a_i}$ that rotates with the jet's -- a priori unknown -- whipping frequency $\Omega$, $\boldsymbol{\Omega}=\Omega \mathbf{a_3}$, $\Omega \in \mathbb{R}$ (see Fig.~\ref{fig:photos}, left). In addition, the director triad for the jet is modified by incorporating the respective spin in order to get stationary boundary conditions \cite{ribe:p:2004}, it becomes
\begin{align*}
\partial_t\mathbf{d}_\mathbf{i}^s=(\boldsymbol{\omega}-u\boldsymbol{\kappa}+\Omega\mathbf{d}_\mathbf{3}^s)\times\mathbf{d}_\mathbf{i}^s,\quad i=1,2,3.
\end{align*}
A representation in these director and outer bases eliminates the time-dependencies and yields a stationary set-up, but it obviously introduces fictitious body forces and couples, such as Coriolis, centrifugal and spin-associated ones, due to inertia in the model equations. The director and outer bases are related by the tensor-valued rotation $\mathbf{R}$, i.e.,\ $\mathbf{R}=\mathbf{a_i}\otimes \mathbf{d}_\mathbf{i}^s$.  For any quantity we use the following coordinate terminology:
\begin{align*}
\mathbf{y}=\sum_{i=1}^3 y_i \mathbf{d}_\mathbf{i}^s =\sum_{i=1}^3 \breve y_i \mathbf{a_i}\in \mathbb{E}^3
\end{align*}
with $\mathsf{y}=(y_1, y_2, y_3)\in \mathbb{R}^3$ and $\mathsf{\breve y}=(\breve{y}_1, \breve{y}_2, \breve{y}_3)\in \mathbb{R}^3$ where $\mathsf{y}=\mathsf{R }\cdot \mathsf{\breve y}$ and $\mathsf{R}=(R_{ij})=(\mathbf{d}_\mathbf{i}^s\cdot \mathbf{a_j})\in SO(3)$. A similar transformation into a stationary set-up has been performed in the investigation of viscous rope coiling \cite{ribe:p:2004}. We emphasize that with this approach we do not artificially insert whipping into the problem. Because the whipping frequency belongs to the solution of the problem, it can also be $\Omega=0$ in certain parameter settings. However, the ansatz for the rotation is only valid if $\mathbf{d_3}(0,t)$ is aligned with $\mathbf{a_3}$ as shown with mathematical arguments in \cite{rivero:p:2015}.

\begin{remark}[Stationarity]\label{rem:stationarity}
The periodic rotation of the system around the symmetry axis $\mathbf{a_3}$ allows alternatively also the following approach to obtain stationarity. We can express any scalar $y$ and vector-valued $\mathbf{y}$ variables as $y(s,t)=y(s,0)$ and $\mathbf{y}(s,t)= \mathbf{M}(t) \cdot \mathbf{y}^\circ(s)$, where 
\begin{align*}
\mathbf{M} (t)= \cos(\Omega t) \mathbf{P}_0(0,0) + \sin (\Omega t) \mathbf{a_3} \times \mathbf{P}_0(0,0) + \mathbf{a_3}\otimes\mathbf{a_3}
\end{align*}
represents the rotation tensor with respect to the jet's whipping frequency $\Omega$ and the fixed reference triad at the nozzle $\mathbf{d}_\mathbf{i}^\circ(0)$  at $t=0$
with $\mathbf{d}_\mathbf{3}^\circ(0)=\mathbf{a_3}$, here
$\mathbf{P}_0(0,0)=\mathbf{d}_\mathbf{1}^\circ(0)\otimes \mathbf{d}_\mathbf{1}^\circ(0)+\mathbf{d}_\mathbf{2}^\circ(0)\otimes \mathbf{d}_\mathbf{2}^\circ(0)$.
Hence, we get $\partial_t y = 0$ and $\partial_t \mathbf{y}(s,0)= \Omega \mathbf{a_3} \times \mathbf{y}^\circ(s)$ and, in particular, $\mathbf{v}^\circ -u^\circ \mathbf{d}_\mathbf{3}^\circ= \Omega \mathbf{a_3} \times \mathbf{r}^\circ$ and $\boldsymbol{\omega}^\circ -u^\circ \boldsymbol{\kappa}^\circ= \Omega (\mathbf{a_3}-\mathbf{d}_3^\circ) $ for the linear and angular velocities. In this consideration the relevant two frames are the director triad $\mathbf{d}_\mathbf{i}^\circ(s)$ and the reference triad $\mathbf{d}_\mathbf{i}^\circ(0)$.
\end{remark}

The derivation of the stationary model equations is analogue to the one for rotational spinning, see \cite{arne:p:2011} for details. In the stationary set-up the flow rate is constant, i.e., $Q=Au=const$, such that $A$ can be replaced by $1/u$ in the equations. In addition, the linear and angular velocities can be expressed in terms of the other variables. Incorporating the viscous material laws leads to a boundary value problem of ordinary differential equations for jet curve $\mathbf{r}$, triad (rotational group) $\mathbf{R}$, curvature $\boldsymbol{\kappa}$, convective speed $u$, contact forces $\mathbf{n}$ and couples $\mathbf{m}$. For the spun jet that leaves the nozzle vertically and straight-lined we have the following geometric and kinematic boundary conditions at the nozzle $(s=0)$ as well as stress-free conditions at the end $(s=L)$
\begin{align*}
\mathbf{r}(0)=\mathbf{0}, \qquad \mathbf{d}_\mathbf{i}^s(0)=\mathbf{a_i}, \qquad \boldsymbol{\kappa}(0)=\mathbf{0}, \qquad u(0)=U, \qquad \mathbf{n}(L)=\mathbf{0}, \qquad \mathbf{m}(L)=\mathbf{0}.
\end{align*}
To determine the further two unknowns of the problem that are the jet length $L$ and the whipping frequency $\Omega$, we impose two additional geometric conditions on the curve's end point. We prescribe its phase and its height in the device geometry
\begin{align*}
 \breve r_1(L) = 0, \qquad \breve r_3(L) = H.
\end{align*}
In spite of the posed condition on the phase, the choice of the outer basis $\{\mathbf{a_1}(t),\mathbf{a_2}(t),\mathbf{a_3}\}$ still leaves one further degree of freedom, e.g.\ the sign of $\breve{r}_2(L)$. Moreover, the sign of $\Omega$ is free. Thus, four stationary solutions are similarly possible in this model framework: $\Omega \geq 0$ or $\Omega \leq0 $ as well as $\breve{r}_2(L) \geq 0$ or $\breve{r}_2(L) \leq 0$. The form of the solutions is invariant, the resulting jets only differ in the rotational direction of the whipping and in the position of the end point with respect to the $\mathbf{a_2}$-halfspace. We aim for the case: $\Omega \geq 0$ and $\breve{r}_2(L)\geq0$. This implies a positive rotation in the outer frame, a positive Rossby number that characterize the relation between the inertial and rotational forces as well as an end point in the positive $\mathbf{a_2}$-halfspace. 

For the numerical treatment it is convenient to deal with the dimensionless model equations that are stated in the director basis. Thus, we make the system dimensionless using the three problem-relevant lengths (jet length $L$, distance between nozzle and counter-electrode $H$, nozzle diameter $D$) and the jet speed at the nozzle $U$. We introduce the dimensionless quantities as $\tilde{\mathsf{y}}(\tilde{s})=\mathsf{y}(\bar{s}\tilde{s})/\bar y$. The reference values are $\bar{s}=L$, $\bar{r}=H$, $\bar{\kappa}=1/H$, $\bar{u}=U$, $\bar{n}=\pi\mu UD^2/(4H)$ and $\bar{m}=\pi\mu UD^4/(16H^2)$. The dimensionless jet length is scaled with the device height, the dimensionless whipping frequency corresponds to the inverse Rossby number, i.e., $\tilde{L} = L/H$ and $\tilde{\Omega} = \Omega H/U$. To keep the notation simple we suppress the label $\tilde{~}$ and also use the symbols $L$ and $\Omega$ for the dimensionless unknowns in the following. The dimensionless stationary model for the jet's whipping regime is then given by
\begin{equation}\label{eq:model}
\begin{aligned}
L^{-1}\mathsf{R}\cdot\partial_s\mathsf{\breve r}&=\mathsf{e_3}, \\
L^{-1}\partial_s\mathsf{R}&= -\mathsf{\kappa} \times \mathsf{R},\\
L^{-1}\partial_s \mathsf{\kappa}&=-\frac{1}{3} \mathsf{\kappa} n_3+\frac{4}{3}u\mathsf{P_{3/2}}\cdot\mathsf{m} 
+\frac{\Omega}{u}\mathsf{\kappa}\times\mathsf{e_3}, \\
L^{-1}\partial_su&=\frac{1}{3}un_3,\\
L^{-1}\partial_s\mathsf{n}
&=-\mathsf{\kappa}\times\mathsf{n}+\mathrm{Re}\, u\left(\mathsf{\kappa}\times\mathsf{e_3} + \frac{1}{3}n_3\mathsf{e_3}\right)
+2 \mathrm{Re}\Omega \left(\mathsf{R}\cdot\mathsf{e_3}\right)\times\mathsf{e_3} \\
&\quad
+\mathrm{Re}\frac{\Omega^2}{u}\mathsf{R}\cdot\left(\mathsf{e_3}\times(\mathsf{e_3}
\times\mathsf{\breve r})\right)
-\mathsf{f}_{ca}-\mathsf{f}_{el} -\mathsf{f}_{air}, \\
L^{-1}\partial_s\mathsf{m}
&=-\mathsf{\kappa}\times\mathsf{m}+\frac{4}{\epsilon^2}\mathsf{n}\times \mathsf{e_3}
+\frac{\mathrm{Re}}{3}\left(u\mathsf{P_3}\cdot\mathsf{m}-\frac{1}{4}n_3\mathsf{P_2}\cdot \mathsf{\kappa}\right) \\
&\quad -\frac{\mathrm{Re}}{4} \frac{\Omega }{u}\mathsf{P_2}\cdot\bigg{(}\frac{1}{3}\mathsf{R}\cdot\mathsf{e_3} n_3- \frac{1}{3}\mathsf{e_3} n_3
+\left(\mathsf{\kappa}-\frac{\Omega}{u}\mathsf{e_3}\right)\times\mathsf{R}\cdot\mathsf{e_3}\bigg{)} \\
& \quad-\frac{\mathrm{Re}}{4} \left(\frac{1}{u^2}\mathsf{P_2}\cdot(u\mathsf{\kappa}-\Omega\mathsf{e_3}+\Omega\mathsf{R}\cdot\mathsf{e_3}) \right) \times \left(u\mathsf{\kappa}-\Omega\mathsf{e_3} +\Omega\mathsf{R}\cdot\mathsf{e_3} \right)
\end{aligned}
\end{equation}
with the capillary, electric and air resistance forces
\begin{align*}
\mathsf{f}_{ca}&=\Gamma\frac{1}{\sqrt{u}}\left(2\kappa\times\mathsf{e_3}-\frac{1}{3}n_3\mathsf{e_3}\right),\\
\mathsf{f}_{el}&=\Xi\left(1-\frac{\Lambda}{4}\frac{1}{u}(\mathsf{R}\cdot \mathsf{e_3})\cdot \mathsf{e_3}\right)\left(4\frac{1}{u}\mathsf{R}\cdot\mathsf{e}_{3}- \Theta\left(1-\frac{\Lambda}{4}\frac{1}{u}(\mathsf{R}\cdot \mathsf{e_3})\cdot \mathsf{e_3}\right)\frac{1}{u^2}\log\left(\frac{2}{\epsilon}\sqrt{u}\right)\kappa\times\mathsf{e_3}\right),\\
 \mathsf{f}_{air} &= \mathrm{M}\mathrm{Re}\sqrt{u}~\mathsf{F}\bigg(\mathsf{e_3},-\mathrm{Re}_\star\frac{1}{\sqrt{u}}(u\mathsf{e_3}+\Omega\mathsf{R}\cdot(\mathsf{e_3}\times\mathsf{\breve{r}}))\bigg)
\end{align*}
and the boundary conditions
\begin{align*}
\mathsf{\breve r}(0) &= \mathsf{0}, &&\mathsf{R}(0) = \mathsf{P}_1, && \kappa(0) = \mathsf{0}, && u(0) = 1 \\
 \breve r_1(1) &= 0, && \breve r_3(1) = 1, && \mathsf{n}(1) = \mathsf{0}, && \mathsf{m}(1) = \mathsf{0},
\end{align*}
where $\mathsf{P}_k=\mathrm{diag}(1,1,k)$, $k\in \mathbb{R}$ and $\{\mathsf{e_1},\mathsf{e_2},\mathsf{e_3}\}$ is the canonical basis in $ \mathbb{R}^3$. Moreover we write the air drag function $\mathbf{F}$ with respect to the spin-associated director triad using $\mathsf{F} = (F_1,F_2,F_3)$ with $\sum_{i=1}^3 F_i(\tau,\mathsf{w}) \mathbf{d}_\mathbf{i}^\mathbf{s} = \textbf{F}\big(\sum_{i=1}^3\tau_i \mathbf{d}_\mathbf{i}^\mathbf{s}, \sum_{i=1}^3 w_i \mathbf{d}_\mathbf{i}^\mathbf{s} \big).$ Then, the jet's whipping model \eqref{eq:model} is characterized by eight dimensionless parameters
\begin{align*}
\mathrm{Re}&=\frac{\rho U H}{\mu}, \qquad \Gamma=\frac{\gamma H}{\mu U D},   \qquad   \Xi=\frac{I\Phi H}{\pi \mu U^2 D^2},  \qquad  \Theta=\frac{I}{\pi\varepsilon_p U\Phi},  \qquad \Lambda= \frac{\pi \lambda \Phi D^2}{I H}, \qquad \epsilon=\frac{D}{H},\\
\mathrm{M}&=\frac{4\mu_\star^2 H}{\pi\rho\rho_\star U^2D^3},\qquad  \mathrm{Re}_\star=\frac{\rho_\star UD}{\mu_\star}
\end{align*}
that are the Reynolds number $\mathrm{Re}$ as ratio between inertial and viscous forces, the surface-tension associated number $\Gamma$ (scaled inverse Capillary number), the potential-current-associated numbers $\Xi$, $\Theta$ and $\Lambda$, the slenderness ratio $\epsilon$ between nozzle diameter and device height as well as the air drag-associated numbers $\mathrm{M}$ and $\mathrm{Re}_\star$. The last is obviously also a Reynolds number but with respect to the airflow quantities. 

The conductive effects are represented by $\Lambda$. This characteristic number shows the limitation of our comparatively simple electric force model that assumes a constant external electric field and the current as a parameter. It must hold $\Lambda/4<1$. Otherwise the acting electric forces at $s=0$ point into the nozzle which is not physical. In the following we neglect the term, i.e., $\Lambda=0$, and hence overestimate the electric forces by assuming pure convection.

\begin{remark}[Quaternions for rotation]
To parameterize the rotation $\mathsf{R}\in SO(3)$  we use unit quaternions, i.e., $\mathsf{q}=(q_0,q_1,q_2,q_3)\in \mathbb{R}^4$ with $\|\mathsf{q}\|=1$. Consider
\begin{align*}
\mathsf{R}(\mathsf{q})=
 \left( \begin{array}{ccc}
q_1^2-q_2^2-q_3^2+q_0^2 & 2 (q_1 q_2 - q_0 q_3) & 2 (q_1 q_3 + q_0 q_2)\\
2 (q_1 q_2 + q_0 q_3)  & -q_1^2+q_2^2-q_3^2+q_0^2 & 2 (q_2 q_3 - q_0 q_1)\\
2 (q_1 q_3 - q_0 q_2) & 2 (q_2 q_3 + q_0 q_1) & -q_1^2-q_2^2+q_3^2+q_0^2 
\end{array} \right), 
\end{align*}
the equations $L^{-1} \partial_s \mathsf{R}=-\mathsf{\kappa} \times \mathsf{R}$ with $\mathsf{R}(0)=\mathsf{P}_1$ are replaced in \eqref{eq:model} by
\begin{align}\label{eq:q}
 L^{-1} \partial_s\mathsf{q}=\mathcal{A}(\mathsf{\kappa})\cdot \mathsf{q}, \quad \mathsf{q}(0)=(1,0,0,0), && \
\mathcal{A}(\mathsf{\kappa}) = \frac{1}{2}
\left( \begin{array}{cccc}
0 & \kappa_1 & \kappa_2 & \kappa_3\\
-\kappa_1 & 0 & \kappa_3 & -\kappa_2\\
-\kappa_2 & -\kappa_3 & 0 & \kappa_1\\
-\kappa_3 & \kappa_2 & -\kappa_1 & 0
\end{array} \right).
\end{align}
\end{remark}

\setcounter{equation}{0} \setcounter{figure}{0}
\section{Continuation-Collocation Method}\label{sec:3}

The derived Cosserat rod model for the jet's whipping \eqref{eq:model}-\eqref{eq:q} is a parametric boundary value problem of ordinary differential equations with 19 variables. The numerical challenge lies in solving the problem for arbitrary parameter settings, which requires suitable initial guesses of the respective solutions. The difficulty of finding a suitable guess was already addressed in \cite{ribe:p:2004} for viscous rope coiling and treated there manually. We present here a continuation-collocation method that makes the efficient and robust simulation and automatic navigation through a high-dimensional parameter space possible.

\subsection{Collocation scheme}
To solve a boundary value problem of the form
\begin{align}\label{eq:bvp}
\frac{\mathrm{d}}{\mathrm{d}s}\mathsf{y}=\mathsf{f}(\mathsf{y}), \qquad \mathsf{g}(\mathsf{y}(0),\mathsf{y}(1))=\mathsf{0}
\end{align} 
we use a four-stage Lobatto IIIa formula as collocation scheme \cite{hairer:b:2009}. It is an implicit Runge-Kutta method. The collocation polynomial provides an once continuously differentiable solution that is fifth-order accurate uniformly in $s \in [0,1]$. Mesh selection and error control are based on a scaled residual and the true error of the continuous solution \cite{kierzenka:p:2008}. Thus, we have
\begin{align*}
&\mathsf{y}_{i+1}-\mathsf{y}_i - h_{i+1}\sum\limits_{j=1}^4 b_j\mathsf{k}_j=\mathsf{0},\qquad \mathsf{g}(\mathsf{y}_0,\mathsf{y}_N)=\mathsf{0}\\
&\text{with } \mathsf{k}_j = \mathsf{f}\bigg(\mathsf{y}_i+h_{i+1}\sum\limits_{k=1}^4 a_{jk}\mathsf{k}_k\bigg)
\end{align*}
with collocation points $0=s_0<s_1<...<s_N=1$, mesh size $h_i=s_i-s_{i-1}$, the abbreviation $\mathsf{y}_i=\mathsf{y}(s_i)$ and the coefficients
\begin{align*}
(a_{jk}) = \frac{1}{120}\begin{pmatrix}
0 & 0 & 0 & 0\\
11+\sqrt{5} & 25-\sqrt{5} & 25-13\sqrt{5} & -1+\sqrt{5}\\
11-\sqrt{5} & 25+13\sqrt{5} & 25+\sqrt{5} & -1-\sqrt{5}\\
10 & 50 & 50 & 10
\end{pmatrix},\qquad
(b_j) = \frac{1}{12}\begin{pmatrix}
1\\5\\5\\1
\end{pmatrix}.
\end{align*}\noindent
The resulting nonlinear system of $N+1$ equations for $(\mathsf{y}_i)_{i=0,...,N}$ is solved using a Newton method with analytically prescribed Jacobian. This is a classical approach that is provided in the software MATLAB by the routine \emph{bvp5c.m} (see www.mathworks.com). Its applicability depends on the convergence of the Newton method that is crucially determined by the initial guess. We aim for adapting the initial guess iteratively by means of a continuation method, solving a sequence of slightly varying boundary value problems. 

\subsection{Continuation procedure}
In the continuation method we embed the boundary value problem of interest \eqref{eq:bvp} into a family of problems by introducing a continuation parameter tuple $\mathsf{c}\in[0,1]^n$
\begin{align*}
\frac{\mathrm{d}}{\mathrm{d}s}\mathsf{y}=\mathsf{\hat f}(\mathsf{y};\mathsf{c}),  \qquad & \mathsf{\hat g}(\mathsf{y}(0),\mathsf{y}(1);\mathsf{c})=\mathsf{0}, && \mathsf{c}\in[0,1]^n\\
\mathsf{\hat f}(\cdot;\mathsf{\underline 1})=\mathsf{f}, \qquad \mathsf{\hat g}(\cdot,\cdot;\mathsf{\underline 1})=\mathsf{g}, \qquad & \mathsf{\hat f}(\cdot;\mathsf{0})=\mathsf{f}_0, \qquad \mathsf{\hat g}(\cdot,\cdot;\mathsf{0})=\mathsf{g}_0.
\end{align*} 
Here, $\mathsf{\underline 1}$ denotes the $n$-dimensional tupel of ones. The functions $\mathsf{f}_0$, $\mathsf{g}_0$ are chosen in such a way that for $\mathsf{c}=\mathsf{0}$ an analytical solution is known. Given this starting solution, we seek for a sequence of parameter tuples $\mathsf{0}=\mathsf{c}_0, \mathsf{c}_1, \ldots, \mathsf{c}_m = \mathsf{\underline 1}$ such that the solution to the respective predecessor boundary value problem provides a good initial guess for the successor. The solution associated to $\mathsf{c}= \mathsf{\underline 1}$ finally belongs to the original system. By help of the continuation parameters certain terms in the ordinary differential equations can be first excluded, then included. Also the boundary conditions can be varied. The choice of the continuation path decides about failure or success because there are not always existing solutions and several meaningful ways. The core of a robust continuation procedure are the step size control and the choice of a continuation path to navigate through a high-dimensional parameter space. 

To explain the used procedures, we consider at first an one-dimensional parameter space $c\in [0,1]$. Proceeding from an initial continuation step size $\Delta c_0$, a boundary value problem is always solved twice by using one full step and two half steps. If the full step requires more Newton iterations than both half steps together or $k_1$-times more collocation points than the second half step, the continuation step is reduced by a factor $k_2$, otherwise it is increased by $k_2$ for the further computation. If the Newton method fails, the step size is reduced by a factor $k_3$ and the computation is repeated. Certainly, it can happen that no solutions exist for $c>c_{crit}\geq 0$, thus the algorithm for the adaptive step size control has a stopping criterion that is based on a minimal step size. In particular, we use $k_1 = 0.1$, $k_2=1.5$, $k_3=10$, $\Delta c_0=10^{-1}$, $\Delta c_{min}=10^{-14}$. 

In a high-dimensional parameter space $\mathsf{c}\in [0,1]^n$, $n\gg1$ there are numerous possible but also impossible continuation paths. We consider an equidistant grid on the parameter space with size $\delta c$ in every dimension and apply a tree search algorithm where the grid points (parameter tuples) are the nodes in the tree. The traversal of the tree is described by a recursion. Because from a given node there is more than one possible next node when changing one parameter, we use a heuristic to order the nodes. We apply a depth-first search to explore a continuation path as far as possible before backtracking when the solution to a boundary value problem cannot be found. The step from one node to the next node is performed by solving the associated boundary value problem either directly or by help of the adaptive step size control described above.
To accelerate the algorithm we use a heuristic that prefers the diagonal navigation through the parameter space and register all visited nodes on a forbidden list. As grid size we take $\delta c=10^{-1}$, implying $10^{n}$ nodes in total.

\subsection{Model-dependent continuation strategy}
In our application we consider a generalized rod model as basis for the continuation. We extend the Cosserat rod equations \eqref{eq:model}-\eqref{eq:q} by introducing continuation parameters  for the viscous, capillary, electric and air forces $c_\mathrm{Re}, c_\Gamma, c_\Xi, c_\Theta, c_\epsilon, c_\mathrm{M} \in [0,1]$. This means we replace the Reynolds number $\mathrm{Re}$ by the term $c_\mathrm{Re}\mathrm{Re}+(1-c_\mathrm{Re})\mathrm{Re}_0$, analogously for $\Gamma$, $\Xi$, $\Theta$, $\epsilon$ and $\mathrm{M}$ with starting values  $\mathrm{Re}_0$, $\Gamma_0$, $\Xi_0$, $\Theta_0$, $\epsilon_0$, $\mathrm{M}_0$. This system is supplemented with the originally posed boundary conditions
\begin{align*}
\mathsf{\breve r}(0) &= \mathsf{0}, \qquad \mathsf{q}(0) = (1,0,0,0), \qquad u(0) = 1, \qquad \breve r_3(1) = 1
\end{align*}
and the following -- slightly modified -- ones
\begin{equation*}
\begin{aligned}
\mathsf{\kappa}(0)-(1-c_1)\mathsf{e_2}&=\mathsf{0}\\
c_2\breve{r}_1(1)-(1-c_2)\breve{r}_2(1)&=\mathsf{0}\\
(1-c_7)[(1-c_3+c_3\Omega)\mathsf{R}(1)\cdot (\mathsf{e_3}\times \mathsf{\breve{r}}(1))+u(1)\mathsf{e_3}-(1-c_4)(\mathsf{e_2}+\mathsf{e_3})] + c_7 \mathsf{n}(1) &=\mathsf{0}\\
(1-c_7)[(1-c_5+c_5\Omega)(\mathsf{R}(1)\cdot\mathsf{e_3}-\mathsf{e_3})+u(1)\kappa(1)+(1-c_6)(\mathsf{e_1}-\mathsf{e_2}+\mathsf{e_3})] + c_7 \mathsf{m}(1) &=\mathsf{0}.
\end{aligned}
\end{equation*}
Consequently, we embed the jet model into a family of boundary value problems in a high-dimensional parameter space, $\mathsf{c}=( c_1,...,c_7, c_\mathrm{Re},c_\Gamma, c_\Xi, c_\Theta, c_\epsilon, c_\mathrm{M})\in [0,1]^{13}$. The choice of the continuation parameters allows the variation of the boundary conditions (jet end with lay-down or without stresses) as well as the exclusion and inclusion of the viscous, capillary, electric and air drag effects that mainly dominate the jet dynamics. We summarize the dimensionless model parameters in the tuple 
\begin{align*}
\mathrm{p} = (\mathrm{Re},\Gamma,\Xi,\Theta,\epsilon,\mathrm{M}).
\end{align*} 
The remaining model parameter $\mathrm{Re}_\star$ is always directly set as desired and not considered in the continuation.

The starting solution for the continuation that belongs to $\mathsf{c}=\mathsf{0}$ is taken from the study on viscous rope coiling \cite{ribe:p:2004}, i.e.,
\begin{equation}\label{eq:ini}
\begin{aligned}
\mathsf{\breve r}(s)=(1-\cos(\pi s/2),0, \sin(\pi s/2), \qquad &\mathsf{q}(s)=(\cos(\pi s/4),0,-\sin(\pi s/4), 0)\\
\mathsf{\kappa}=(0,1,0), \qquad u=1, \qquad &\mathsf{n}=\mathsf{m}=\mathsf{0}, \qquad L=\pi/2, \qquad \Omega=0. 
\end{aligned}
\end{equation}
It describes a (non-coiling) jet having the form of a quarter circle in the absence of inertia, surface tension and outer forces, $\mathrm{Re}_0=\Gamma_0=\Xi_0=\mathrm{M}_0=0$. The initialization is consistent to our aim $\Omega\geq 0$ and $\breve r_2(1)\geq 0$. The corresponding parameter tuple is denoted by $\mathrm{p}_0 = (\mathrm{Re}_0,\Gamma_0,\Xi_0,\Theta_0,\epsilon_0,\mathrm{M}_0)$ with
\begin{align*}
0 \leq \Theta_0 = \bar{\Theta} \leq \Theta,\quad \epsilon \leq \epsilon_0 = \bar{\epsilon}
\end{align*}
using intermediate parameters $\bar{\Theta}$, $\bar{\epsilon}$. In addition we introduce the tuple $\mathrm{\bar{p}}$ of intermediate parameters that allow for moderate physical effects,
\begin{align*}
\mathrm{Re}_0 \leq \mathrm{\bar{Re}} \leq \mathrm{Re},\qquad \Gamma_0 \leq \bar{\Gamma} \leq \Gamma,\qquad \Xi_0 \leq \bar{\Xi} \leq \Xi,\qquad \mathrm{M}_0 \leq \mathrm{\bar{M}} \leq \mathrm{M}.
\end{align*}
In particular we choose $\bar{\mathrm{p}} = (0.2, 4, 2000, 400, 0.1, 0)$. The starting solution \eqref{eq:ini} satisfies the desired stress-free condition $\mathsf{n}(1)=\mathsf{m}(1)=\mathsf{0}$ that belongs to $c_7=1$. However, the change of each continuation parameter has different effects on the form of the solution and hence on finding appropriate initial guesses in the continuation. To navigate through the high-dimensional parameter space from $\mathsf{c}=\mathsf{0}$ to $\mathsf{c}=\mathsf{\underline 1}$ we follow therefore a strategy that consists of four parts:
\begin{enumerate}
\item[(A)] from $\mathsf{c}=\mathsf{0}$ to $\mathsf{c}^A=(1,1,1,1,1,1,0,0,0,0,0,0,0)$: \\
By changing the boundary conditions we obtain the solution associated to viscous rope coiling with dimensionless model parameters $\mathrm{p}_0$ and vanishing linear and angular velocities at the jet end, $c_i=1$ for $i=1,...,6$.
\item[(B)] from $\mathsf{c}^A$ to $\mathsf{c}^B=(1,1,1,1,1,1,0,c_{\mathrm{\bar{Re}}},c_{\bar{\Gamma}},c_{\bar{\Xi}},
0,0,c_{\mathrm{\bar{M}}})$:\\
By increasing the parameters $c_\mathrm{Re}$, $c_\Gamma$, $c_\Xi$, $c_\mathrm{M}$ to intermediate values we incorporate moderate forming of the viscous, capillary, electric and air drag effects and achieve a solution corresponding to the parameter tuple $\mathrm{\bar{p}}$.
\item[(C)] from $\mathsf{c}^B$ to  $\mathsf{c}^C=(1,1,1,1,1,1,1,c_{\mathrm{\bar{Re}}},c_{\bar{\Gamma}},c_{\bar{\Xi}},
0,0,c_{\mathrm{\bar{M}}})$:\\
By changing $c_7$ we switch from lay-down to stress-free boundary conditions.
\item[(D)] from $\mathsf{c}^C$ to $\mathsf{c} = \mathsf{\underline{1}}$:\\
By the final increase of the parameters $c_\mathrm{Re}$, $c_\Gamma$, $c_\Xi$, $c_\Theta$, $c_\epsilon$, $c_\mathrm{M}$ we incorporate all inertial, capillary, electric and aerodynamic effects as desired.
\end{enumerate} 
As mentioned we use a heuristic that seeks for the diagonal continuation path through the parameter space. In Part A we use the tree search algorithm on a grid with $10^6$ nodes and compute each step directly without refinement. In Part B it turns out that balancing the physical effects strictly leads to the diagonal path such that the parameter space can be reduced to $c_\mathrm{Re}=c_\Gamma=c_\Xi=c_\mathrm{M}$. The corresponding one-dimensional search is performed with the adaptive step size control, the same holds for Part C. In Part D we also use the one-dimensional search by increasing every parameter to the target value successively. A good heuristic is to start with the parameter $c_\Theta$ and continue increasing the remaining parameters in the fixed order $c_\Gamma$, $c_\Xi$, $c_\mathrm{Re}$, $c_\mathrm{M}$, $c_\epsilon$.

\begin{figure}[b]
\includegraphics[width=1.0\textwidth]{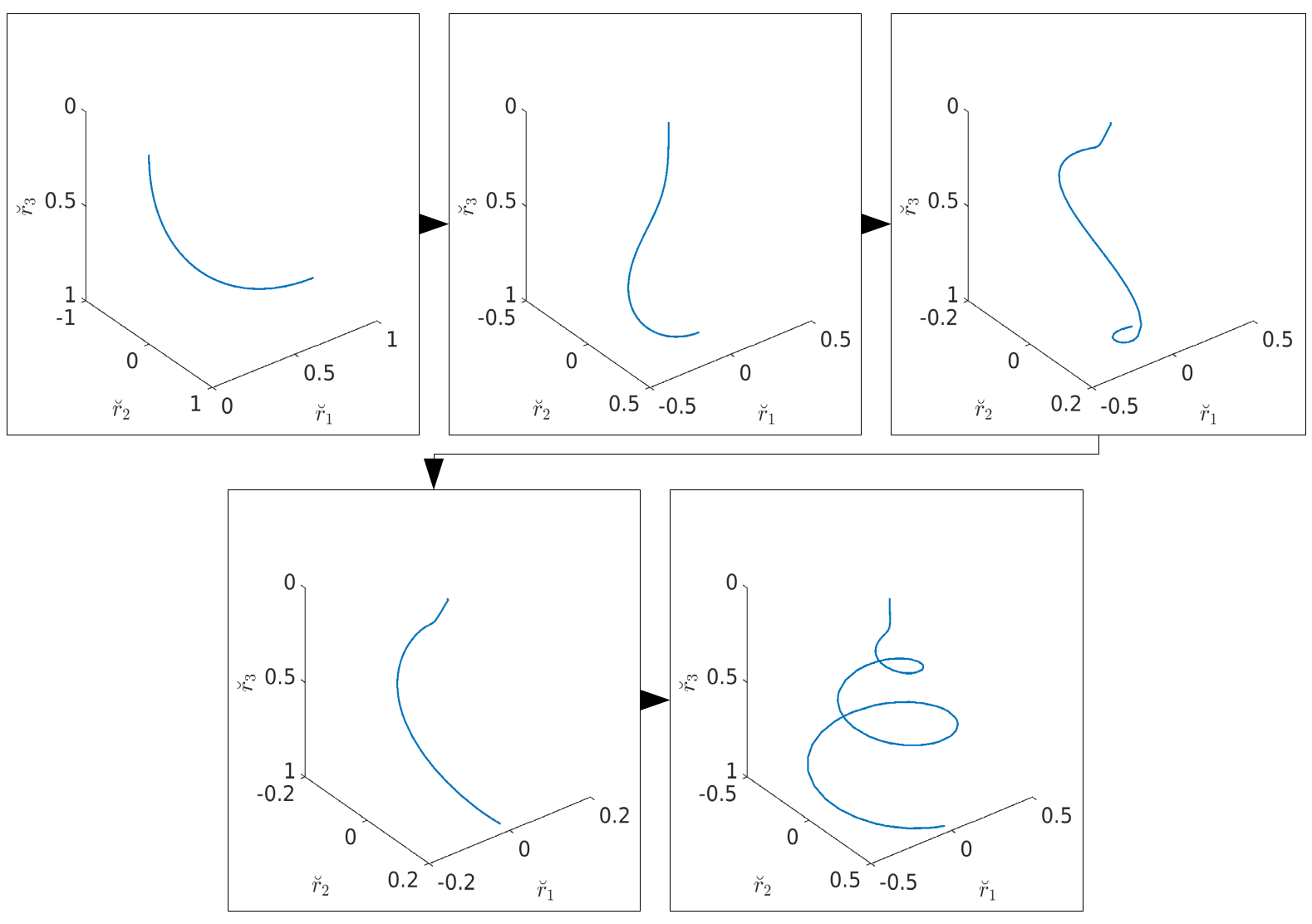}
\caption{Jet curve during continuation procedure. \emph{From top left to bottom right}: Starting solution to $\mathrm{p}_0$, after Part A, after Part B (solution with lay-down end to $\mathrm{\bar p}$), after Part C and after Part D (desired solution with stress-free end to $\mathrm{p}^{ref}$). $\mathrm{Re}_\star=1.5$.
\label{fig:continSteps}}
\end{figure}
\begin{table}[b]
\begin{center}
\begin{tabular}{c | c c | c}\hline
Part & Computing Time [s] & Percentage & Continuation Steps\\ \hline\hline
A & \,\, 173.64 & \,\, 8.6\% & \,\, 61\\
B & \,\, 660.14 & 32.6\% & 136\\
C & \,\, 190.96 & \,\, 9.4\% & \,\, 25\\
D & 1000.04 & 49.4\% & 112\\ \hline
total & 2024.78 & 100\% & 334\\[1ex]
\end{tabular}
\caption{Performance of continuation procedure. Computing time of each Part~A-D and the number of respective continuation steps to reach the reference parameter tuple $\mathrm{p}^{ref}$ (cf.\ Fig.~\ref{fig:continSteps}).}\label{tab:compTime}
\end{center}
\end{table}

\begin{remark}[Selective preconditioning] 
Due to boundary layers occurring in the solution to the variables $\mathsf{n}$ and $\mathsf{m}$ at the nozzle, which we analyze later, the system of ordinary differential equations (\ref{eq:model}) is badly scaled with respect to these quantities. Therefore our numerical approach includes scaling of these variables using the component-wise maximum norm. This means we solve the system (\ref{eq:model}) with variables $n_i$ and $m_i$ ($i=1,2,3$) that are scaled equal to one. This procedure equals a selective preconditioning. 
\end{remark}

\subsection{Performance of continuation algorithm}

The forthcoming numerical simulations are performed on an Intel Xeon X5675 CPU (6 cores, 12 threads) and 98 GBytes of RAM. The collocation-continuation algorithm is implemented in MATLAB version R2015b and the solver \emph{bvp5c.m} is mainly used with the default values. The only modification is that we update the Jacobian every Newton step. 

As an example for the performance study of the continuation algorithm we choose a parameter tuple inducing moderate emergence of all involved effects in the numerical solution. In particular we set
\begin{align*}
 \mathrm{p}^{ref}=(1.2,250,4\cdot 10^4,65,7\cdot 10^{-2},10^{-3}),\qquad \mathrm{Re}_\star^{ref} = 1.5.
\end{align*}
Figure \ref{fig:continSteps} shows the jet curve during the continuation procedure. We initialize with the prescribed solution \eqref{eq:ini} forming a quarter circle without any inertia, surface tension and outer forces. After changing the boundary conditions in continuation Part A the jet shows a coiling onto the lay-down plane. Incorporating viscous, capillary and outer forces in Part B causes the onset of the whipping mode due to the Coulomb interactions, whereas the external electric field straightens the jet. After Part C we clearly see a stress-free jet end. Finally, the inclusion of all effects in their full magnitude gives rise to the whipping of the jet after Part D. Considering the computational performance of the algorithm (Tab.~\ref{tab:compTime}) we see that the preparatory Part A takes a CPU time of 174 sec. Each step (including rejection) requires about 2-4 sec. In view of the huge tree with $10^6$ nodes and the fact that the continuation path is by no means the diagonal, this is a spectacularly good performance. Part C is in general not very sensitive: approx.\ 191 sec.\ CPU time, the minimal step size is of the order $\min \Delta c\sim \mathcal{O}(10^{-3})$. Part B and D contain the actual continuation and need the highest effort (together 82\% of the total computation time). These parts take several minutes, $\min \Delta c\sim \mathcal{O}(10^{-8})$. However, with only around hundred steps in total each (instead of $10^8$ for the respective fixed step size), it clearly stresses the efficiency of the presented adaptive step size control.

\setcounter{equation}{0} \setcounter{figure}{0}
\section{Results}\label{sec:4}

Employing our model framework we perform a parameter study. Special attention is paid to the numerical simulation of the whipping of very thin jets and of jets exposed to high aerodynamic drag, as it is typical for electrospinning applications.

\subsection{Influence of dimensionless model parameters}

\begin{figure}[t]
\includegraphics[width=1.0\textwidth]{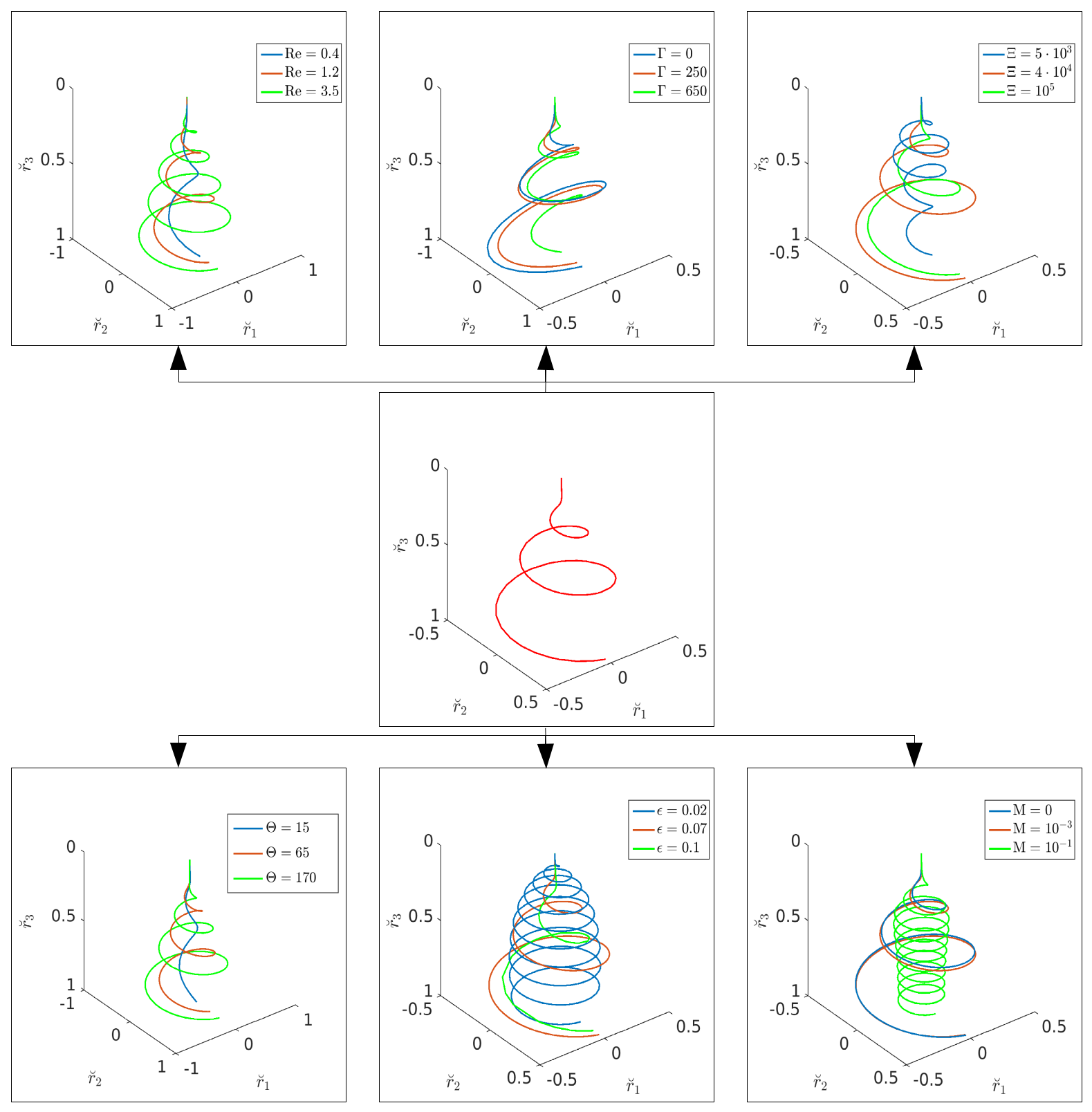}
\caption{Jet curves for variations of single model parameters proceeding from reference $\mathrm{p}^{ref}$, $\mathrm{Re}_\star^{ref}$. The reference jet is visualized in the middle picture and in red in all plots. The jets associated to increased/decreased parameters are given in green/blue.\label{fig:paraVar}}
\end{figure}
\begin{figure}[!t]
\begin{minipage}[t]{0.32\textwidth}
\begin{center}
\includegraphics[width=1.15\textwidth]{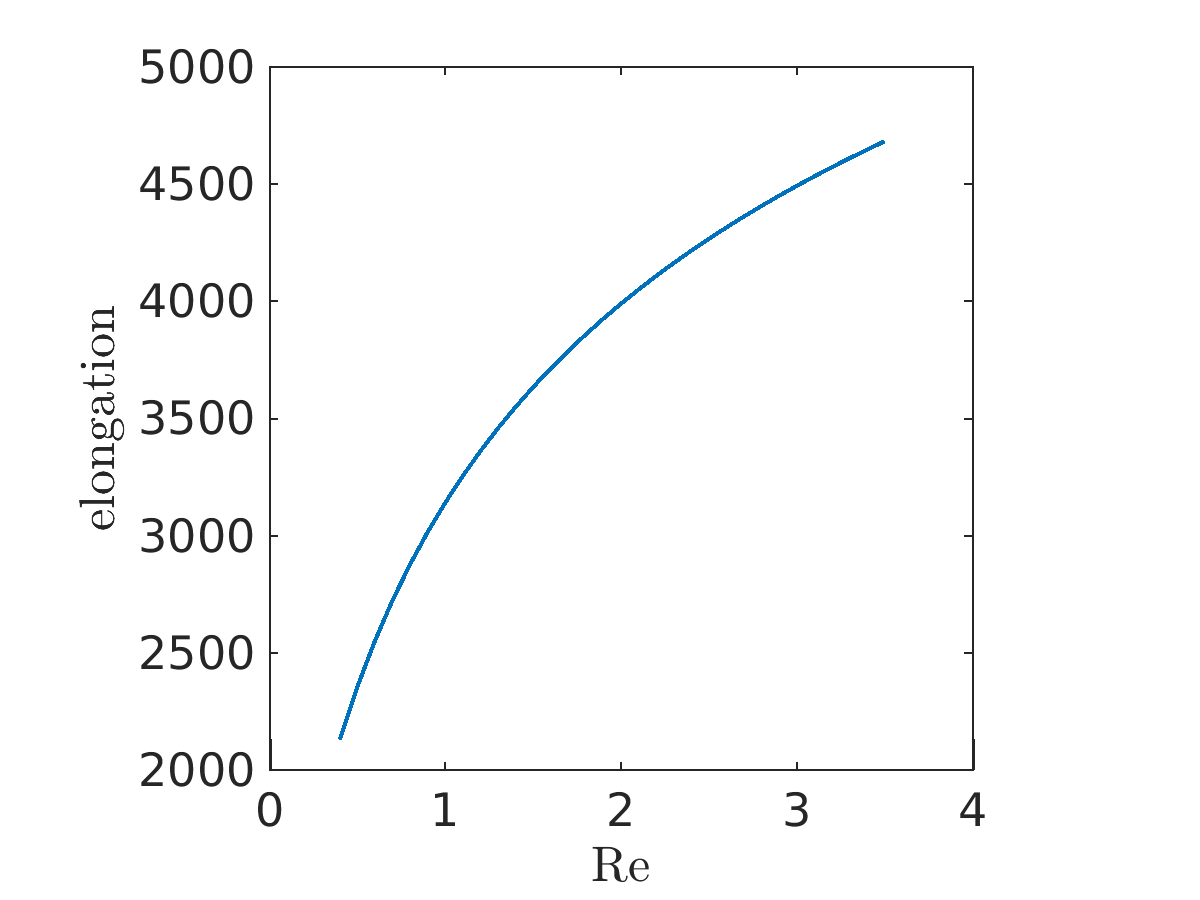}
\end{center}
\end{minipage}
\hfill
\begin{minipage}[t]{0.32\textwidth}
\begin{center}
\includegraphics[width=1.15\textwidth]{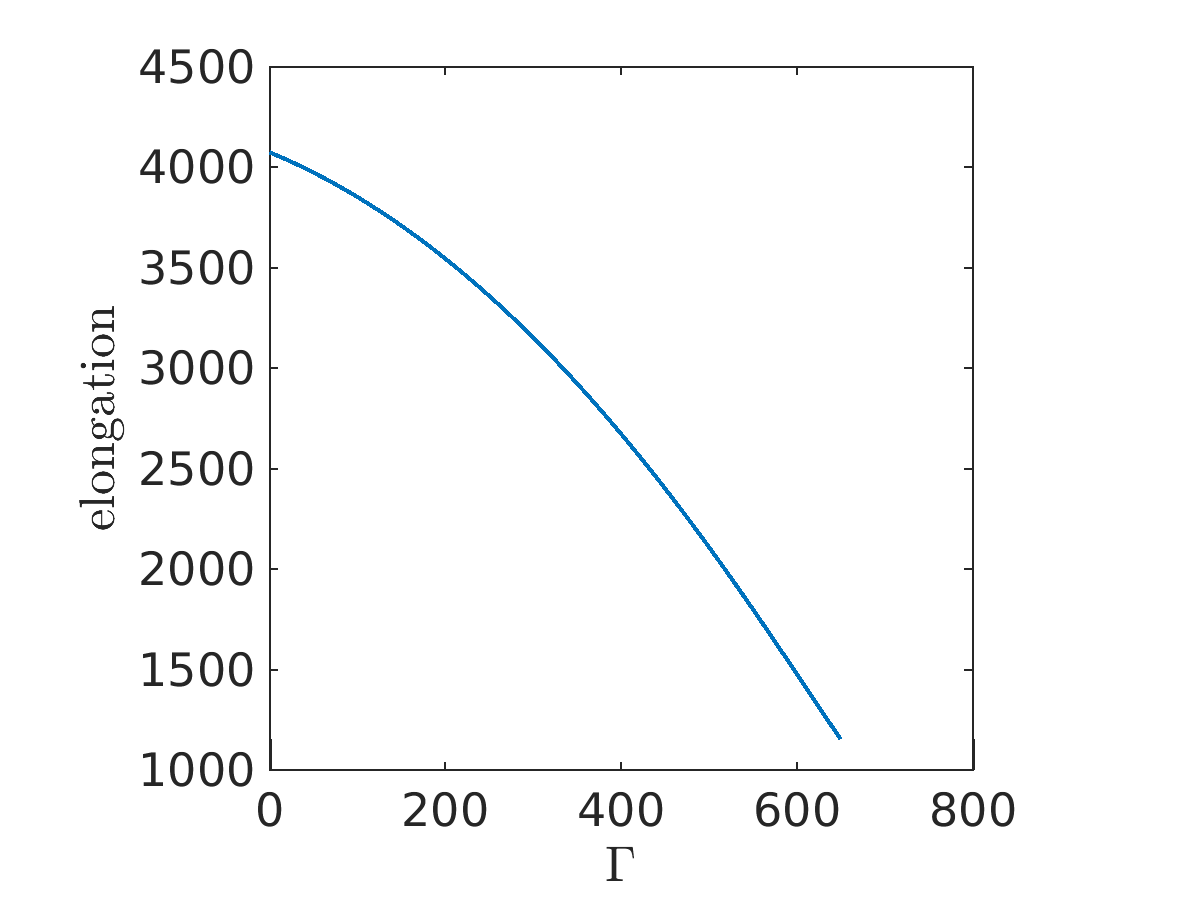}
\end{center}
\end{minipage}
\hfill
\begin{minipage}[t]{0.32\textwidth}
\begin{center}
\includegraphics[width=1.15\textwidth]{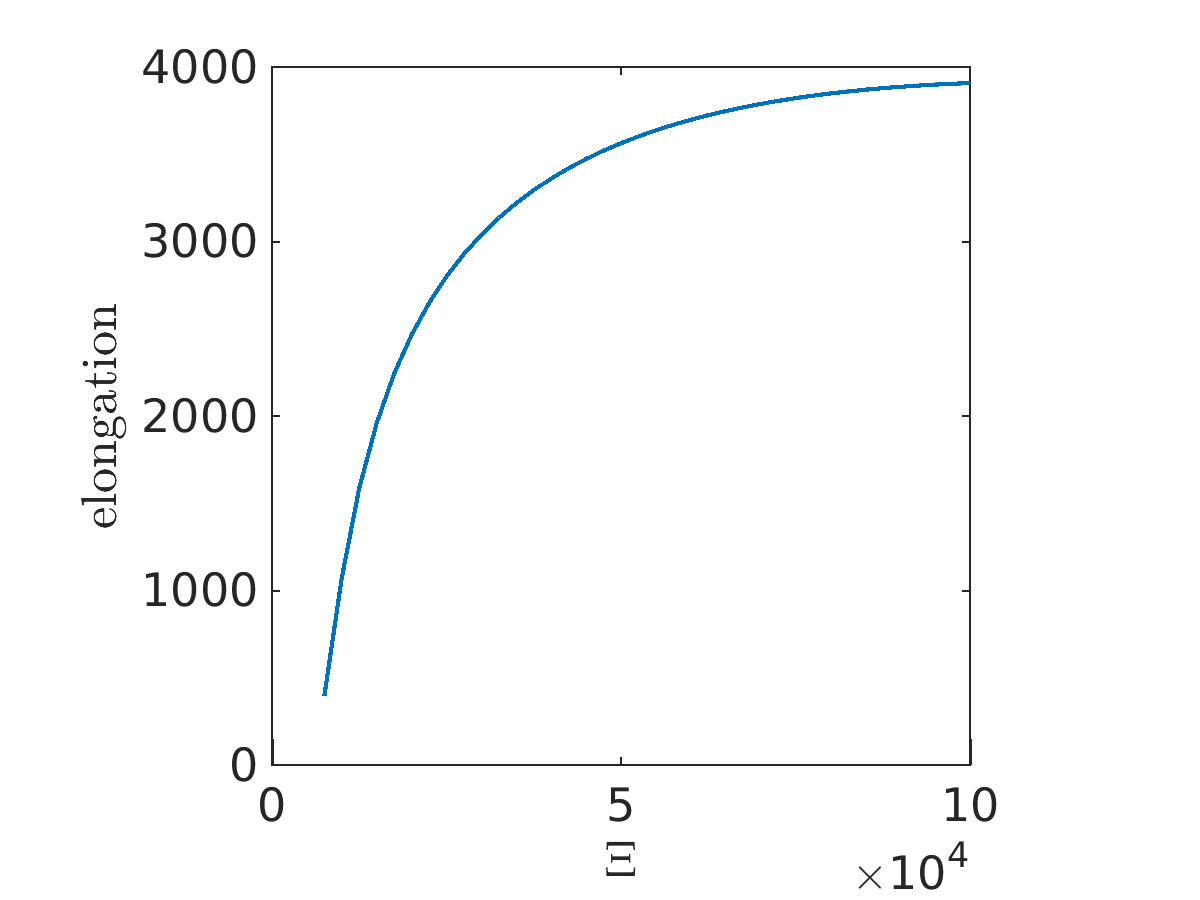}
\end{center}
\end{minipage}
\vfill
\begin{minipage}[t]{0.32\textwidth}
\begin{center}
\includegraphics[width=1.15\textwidth]{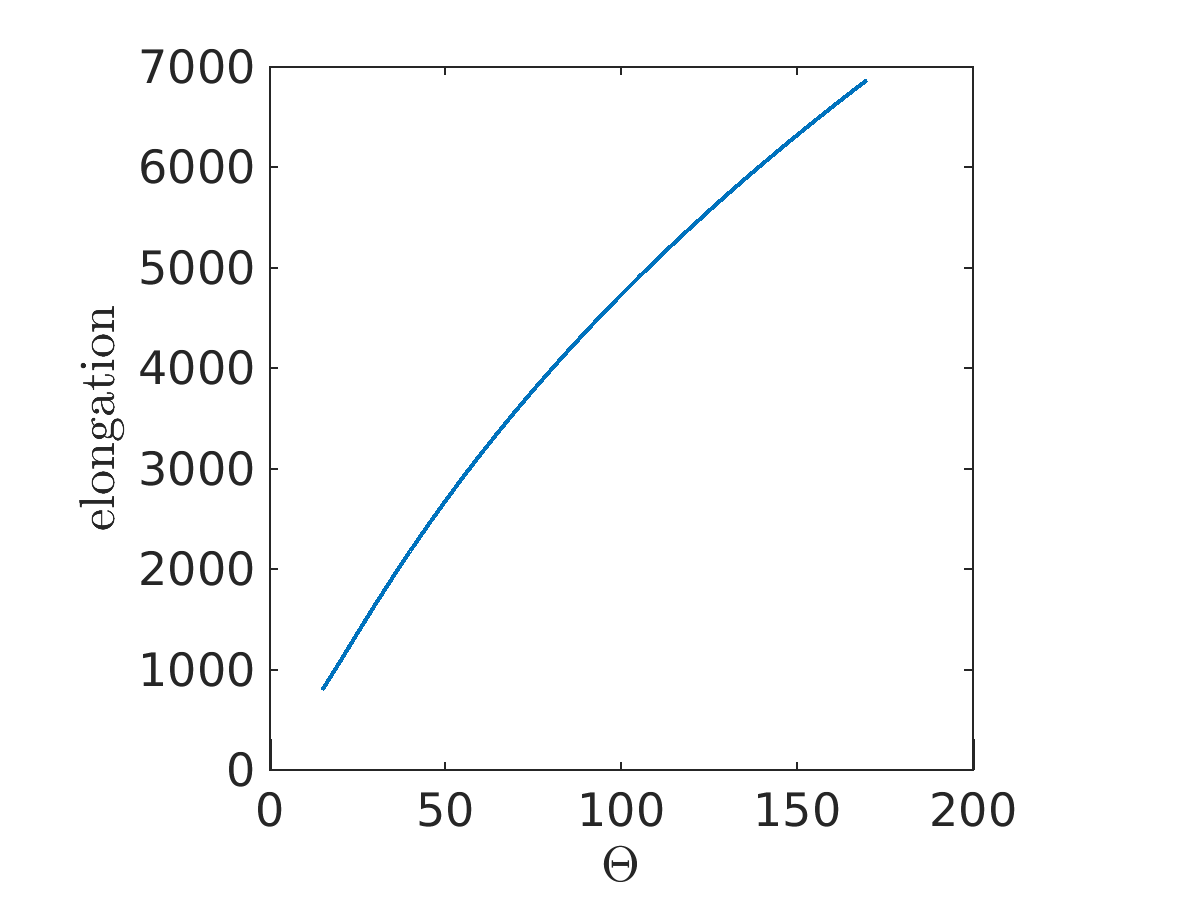}
\end{center}
\end{minipage}
\hfill
\begin{minipage}[t]{0.32\textwidth}
\begin{center}
\includegraphics[width=1.15\textwidth]{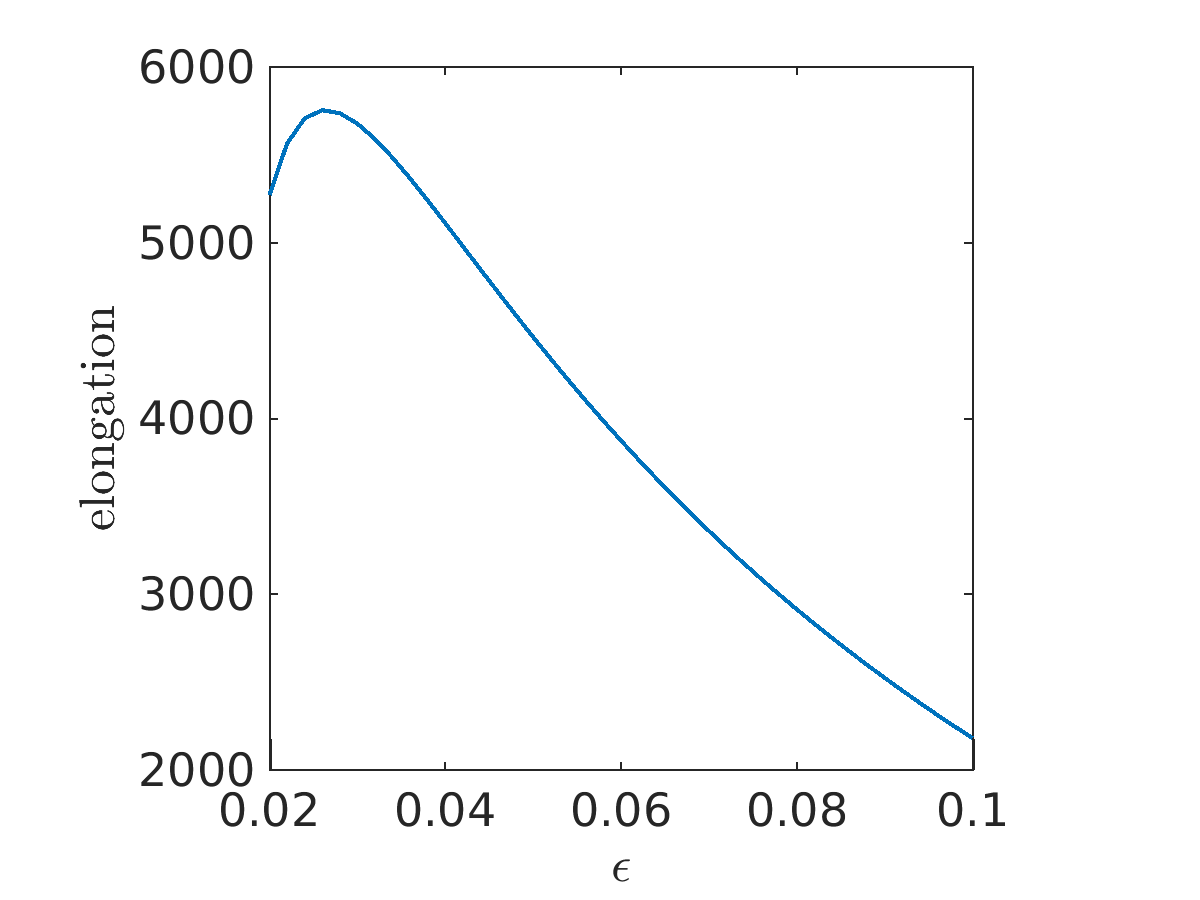}
\end{center}
\end{minipage}
\hfill
\begin{minipage}[t]{0.32\textwidth}
\begin{center}
\includegraphics[width=1.15\textwidth]{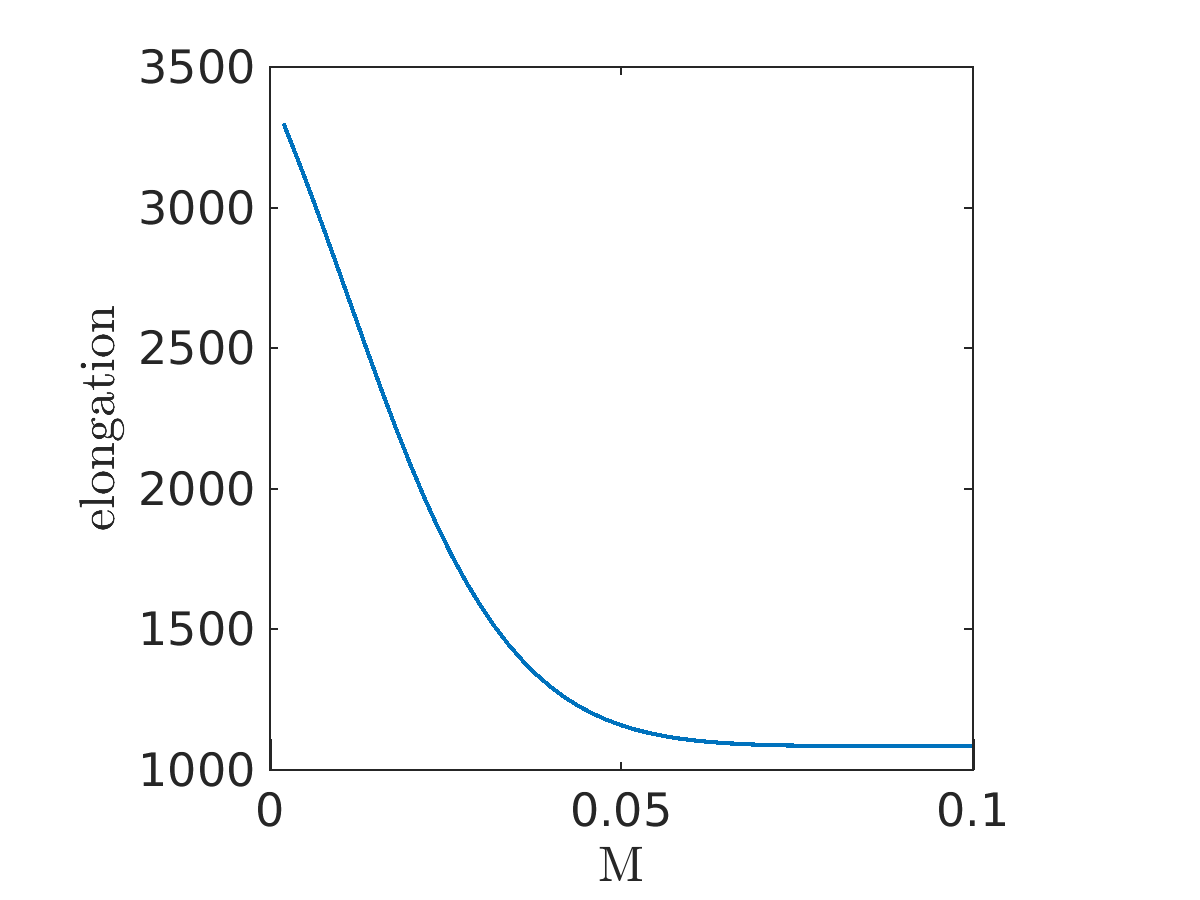}
\end{center}
\end{minipage}
\caption{\label{fig:paraElong}Elongation of jet end over model parameters in the region $\mathrm{Re}\in [0.4,3.5]$, $\Gamma \in [0,650]$, $\Xi\in [5\cdot 10^3,10^5]$, $\Theta\in [15,170]$, $\epsilon\in [0.02,0.1]$, $\mathrm{M}\in [0,0.1]$, proceeding from $\mathrm{p}^{ref}$.}
\end{figure}
\begin{figure}[!t]
\begin{minipage}[t]{0.32\textwidth}
\begin{center}
\includegraphics[width=1.15\textwidth]{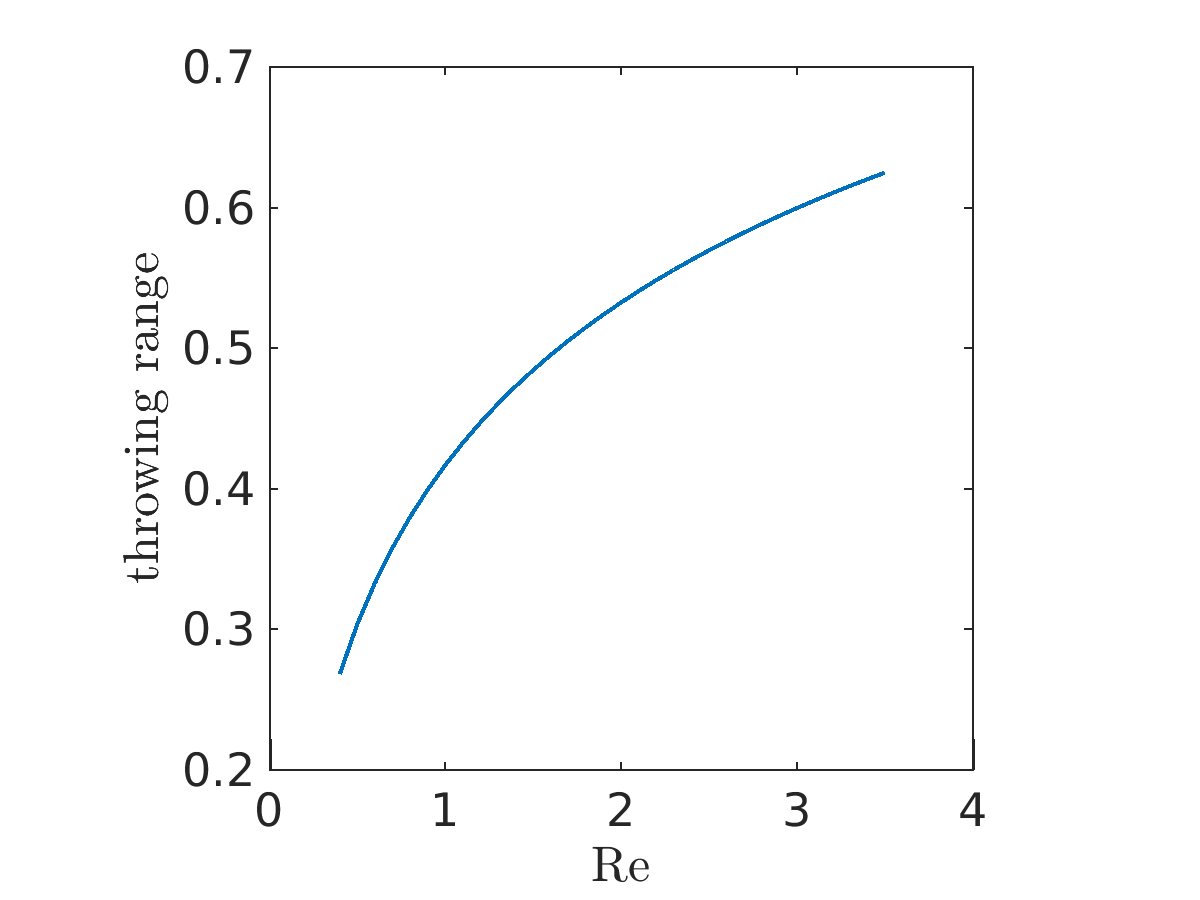}
\end{center}
\end{minipage}
\hfill
\begin{minipage}[t]{0.32\textwidth}
\begin{center}
\includegraphics[width=1.15\textwidth]{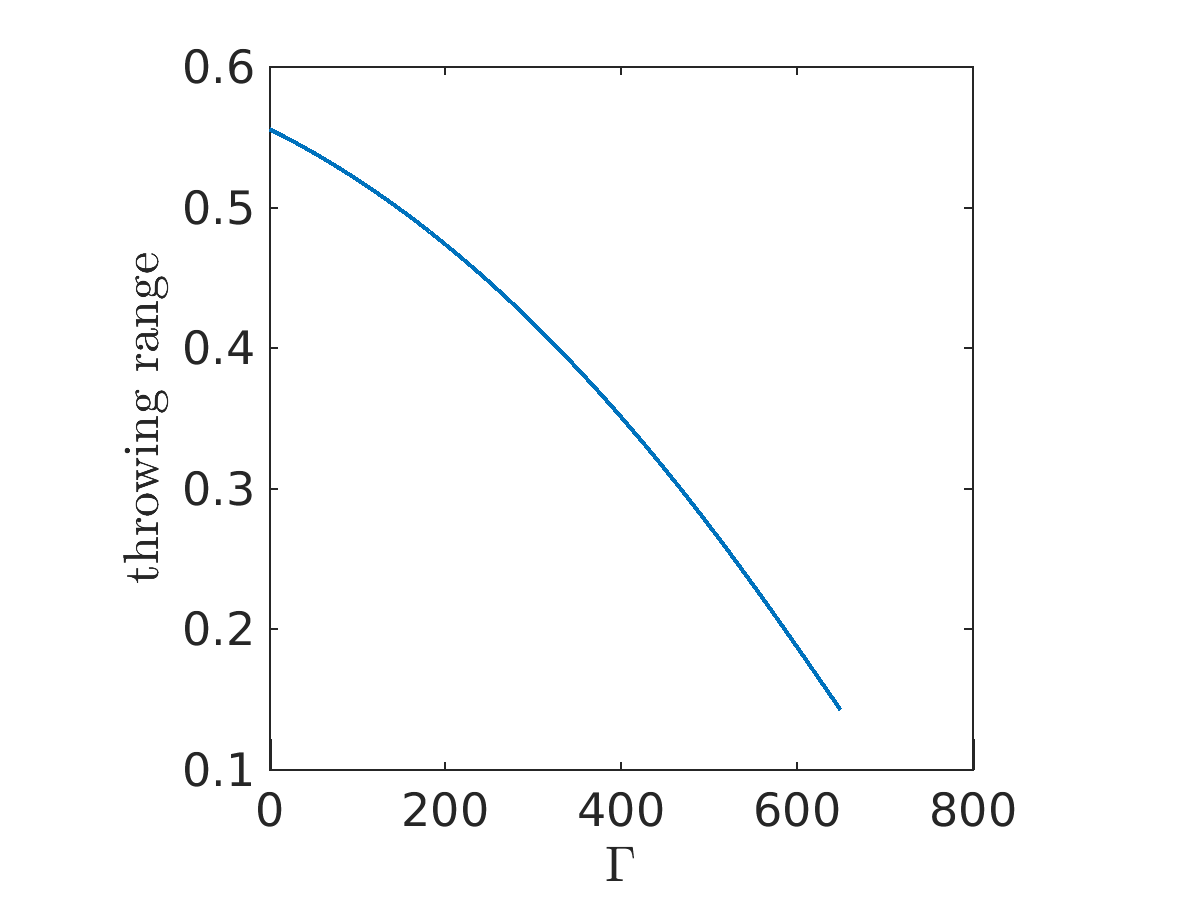}
\end{center}
\end{minipage}
\hfill
\begin{minipage}[t]{0.32\textwidth}
\begin{center}
\includegraphics[width=1.15\textwidth]{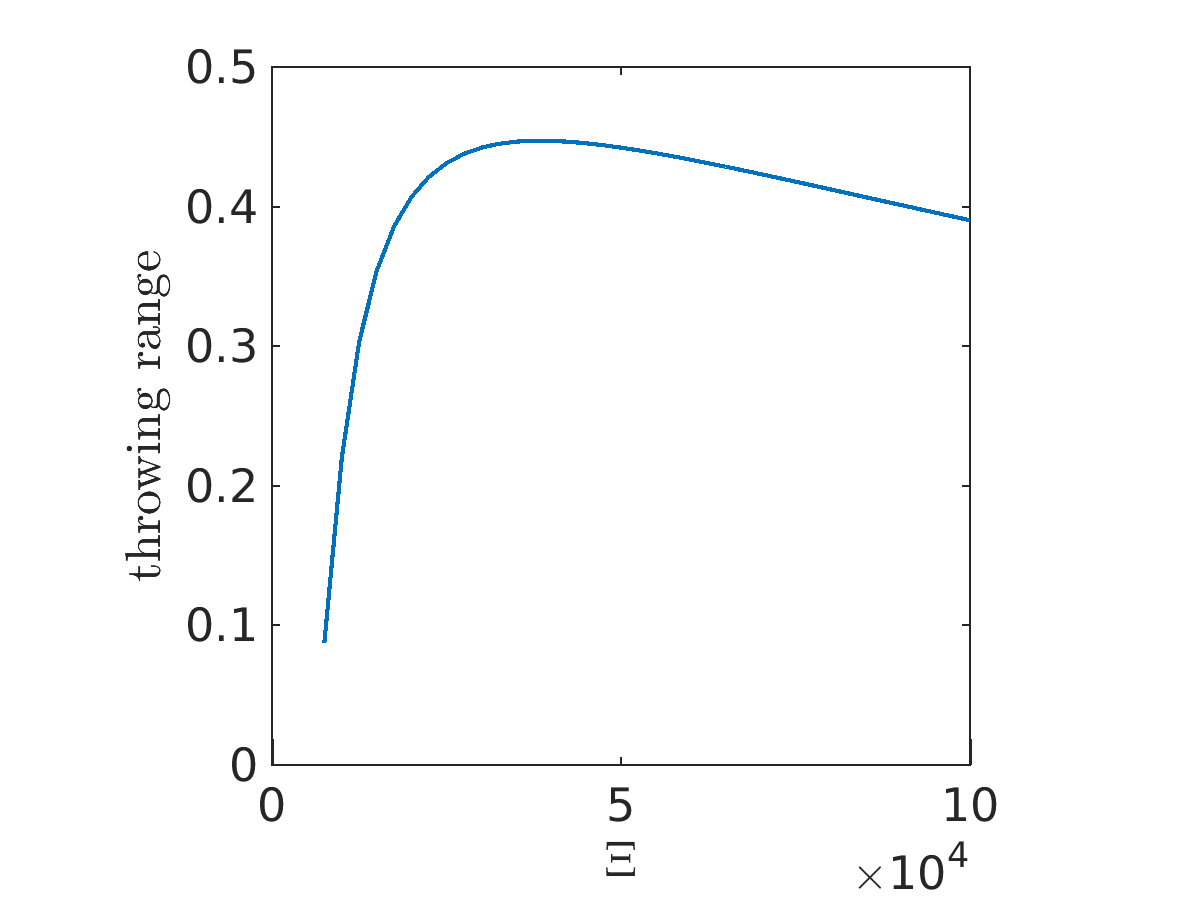}
\end{center}
\end{minipage}
\vfill
\begin{minipage}[t]{0.32\textwidth}
\begin{center}
\includegraphics[width=1.15\textwidth]{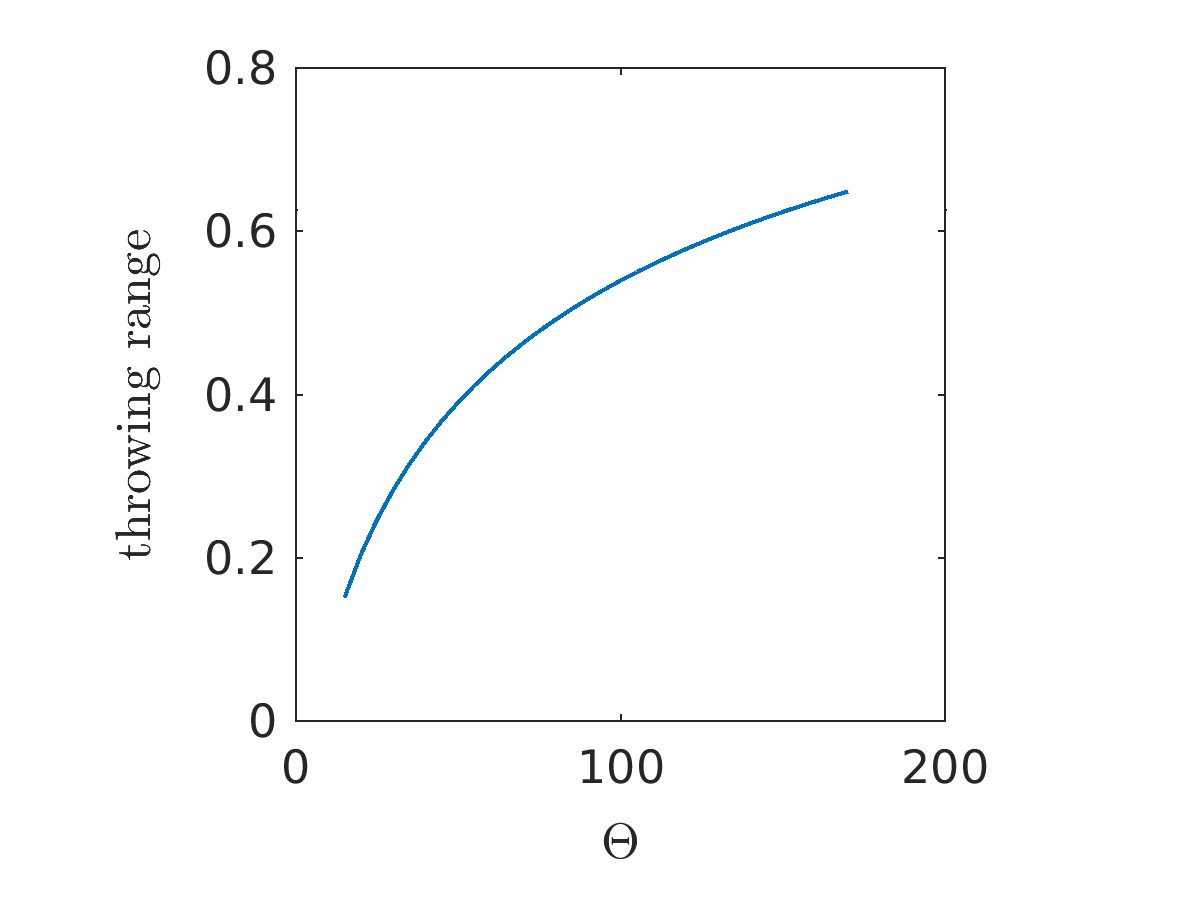}
\end{center}
\end{minipage}
\hfill
\begin{minipage}[t]{0.32\textwidth}
\begin{center}
\includegraphics[width=1.15\textwidth]{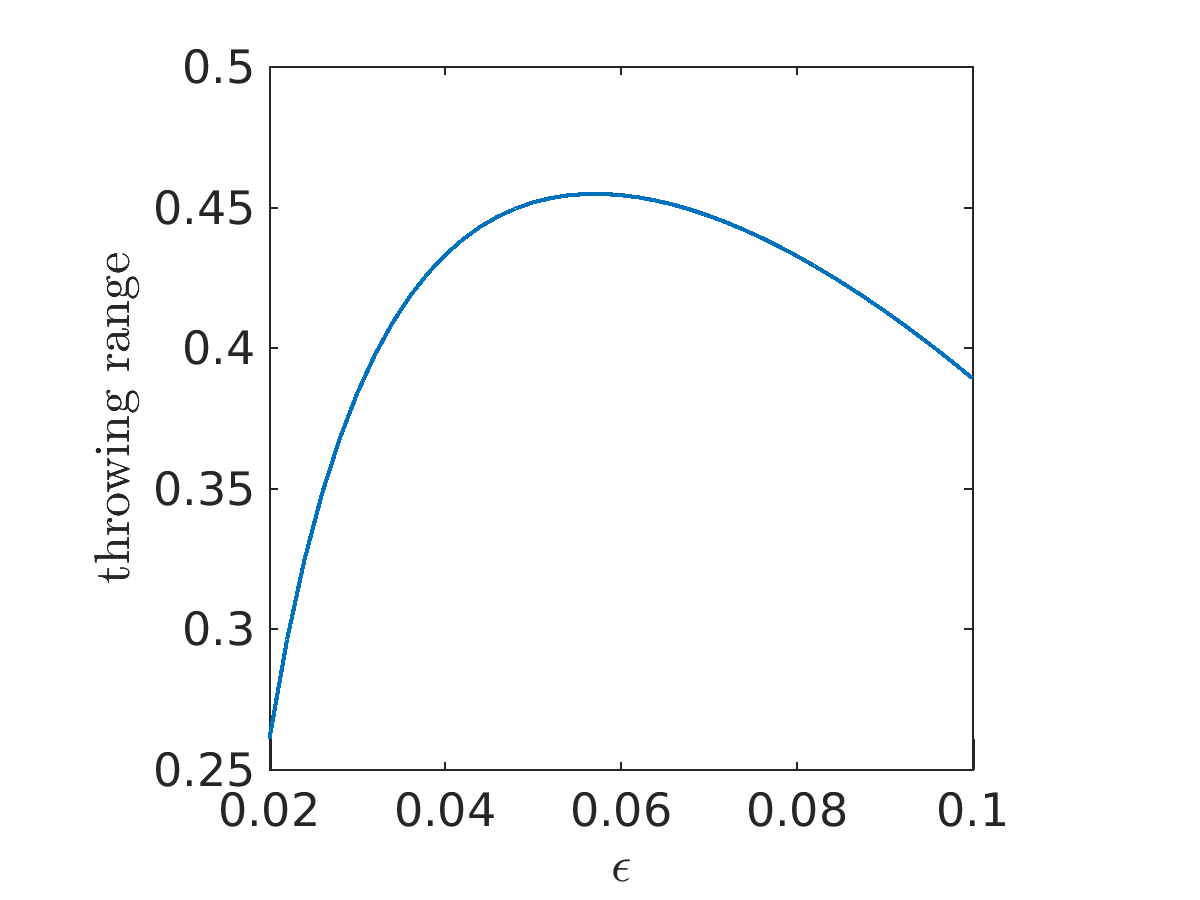}
\end{center}
\end{minipage}
\hfill
\begin{minipage}[t]{0.32\textwidth}
\begin{center}
\includegraphics[width=1.15\textwidth]{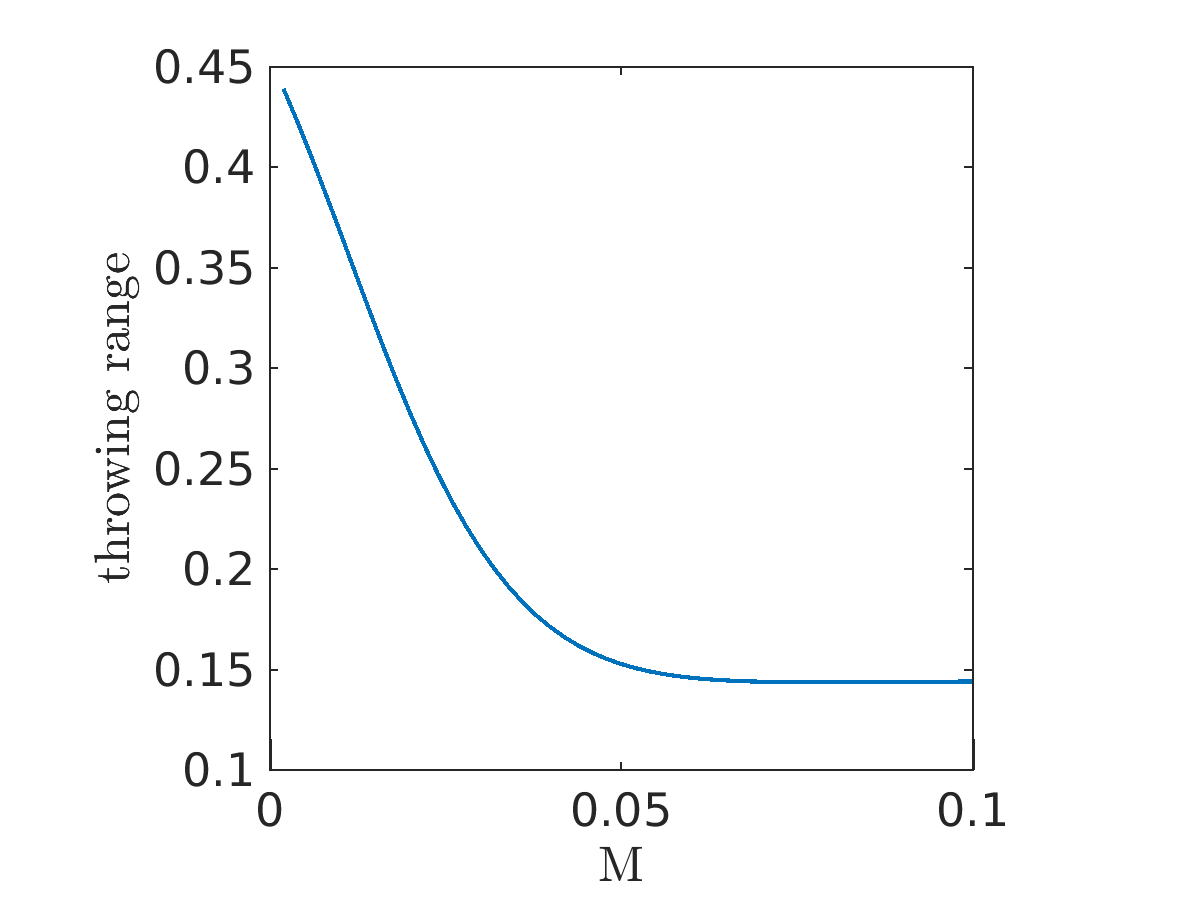}
\end{center}
\end{minipage}
\caption{\label{fig:paraWidth}Jet's throwing range indicating the width of the curve's envelope with respect to the model parameters in the region $\mathrm{Re}\in [0.4,3.5]$, $\Gamma \in [0,650]$, $\Xi\in [5\cdot 10^3,10^5]$, $\Theta\in [15,170]$, $\epsilon\in [0.02,0.1]$, $\mathrm{M}\in [0,0.1]$, proceeding from $\mathrm{p}^{ref}$ (cf.\ Fig.~\ref{fig:paraElong}).}
\end{figure}
Proceeding from the reference solution to $\mathrm{p}^{ref}$ we investigate the jet characteristics under variation of the single model parameters. We particularly study the jet curve, the elongation of the jet end and the throwing range, see Figs.~\ref{fig:paraVar}, \ref{fig:paraElong} and \ref{fig:paraWidth} respectively. Thereby the elongation is given by the Euclidean norm of the jet's tangent in material (Lagrangian) description at the jet end, which corresponds to the convective speed $u(1)$. The throwing range characterizes the width of the envelope enclosing the three-dimensional jet curve and is defined as the distance from the jet's end to the vertical axis, i.e., $\breve{r}_2(1)$.

As we see, an increase of the Reynolds number $\mathrm{Re}$ induces a higher number of turning points coming along with higher elongation and higher throwing range, which can be explained as follows: Increasing $\mathrm{Re}$ equals decreasing the viscous forces under constant inertia. This reduces the thinning of the jet and thus leads to a larger jet surface. As a result we expect a higher surface charge, which comes along with higher Coulomb repulsion inducing a more intense bending profile including higher elongations and throwing ranges. In contrast an increase of the inverse capillary number $\Gamma$ reduces the elongation and the throwing range of the jet including a complete narrowing of the jet's envelope. This takes place due to increasing surface tension forces, which form a counteracting force to the Coulomb repulsion. Furthermore we see that increasing the parameter $\Xi$ induces a more uniform bending due to the stabilizing effect of the outer electric field coming along with a decrease of the Coulomb repulsion. This stabilization procedure is indicated by the formation of a maximal throwing range in the middle of the tested parameter region. Furthermore the elongation reaches an asymptotic bound for large $\Xi$. Varying the parameter $\Theta$ simply means increasing or decreasing the effect of Coulomb repulsion. The effect of stronger Coulomb repulsion is indicated by a higher number of turning points, larger elongation and wider throwing range. Decreasing the typical length scale associated parameter $\epsilon$ leads to a intense increase of turning points. This is clear because following the definition a decrease of this parameter equals an increase of the distance from the nozzle to the ground collector while holding the diameter of the nozzle constant. Therefore the bending jet has more space in $\mathbf{a}_\mathbf{3}$-direction to propagate and thus forms more turning points in total. This higher number of turning points leads to a higher elongation of the jet's end, which reaches a maximum at $\epsilon = 2.6\cdot 10^{-2}$ approximately. The throwing range forms a maximum for values in the middle of the chosen parameter region around $\epsilon = 5.8\cdot 10^{-2}$. Because an increase of $\mathrm{Re}_\star$ simply means a higher air drag acting on the jet and the analogon for decrease, we examine the effect of the surrounding air by varying the parameter $\mathrm{M}$ only. Increasing $\mathrm{M}$ narrows the envelope of the jet considerably and leads to a extensive increase of number of turning points. Considering the curve for the highest value of $\mathrm{M}$ the jet seems to form a helix after a short transient effect. A higher air drag acts as higher counteracting force to the Coulomb repulsion and thus narrows the envelope, which becomes cylindrical as soon as air drag and Coulomb repulsion balance. This effect is also indicated by the rapid drop of elongation and throw range for large air drag parameters until they reach an asymptotic bound.

\subsection{Distinct whipping} 

In electrospinning we are faced with parameter settings involving high whipping frequencies (high numbers of turning points in the jet curve) and large elongations at the jet end. Our stationary model and the proposed numerical scheme are able to handle such situations. However, the required resolution is often computationally very demanding. Hence we make use of the numerical results to formulate an analytical solution for the jet's 'longtime' behavior.

\begin{figure}[b]
\includegraphics[width=0.49\textwidth]{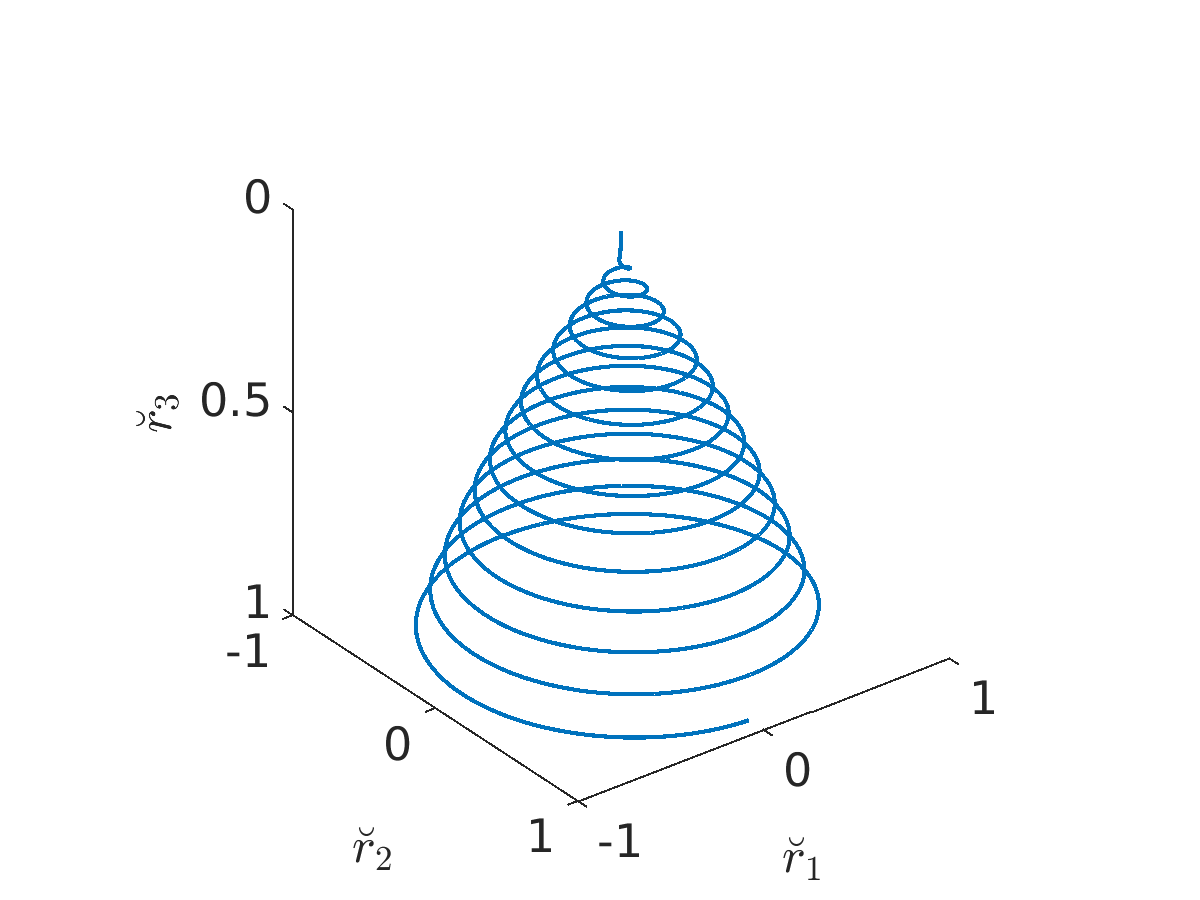}\hfill
\includegraphics[width=0.49\textwidth]{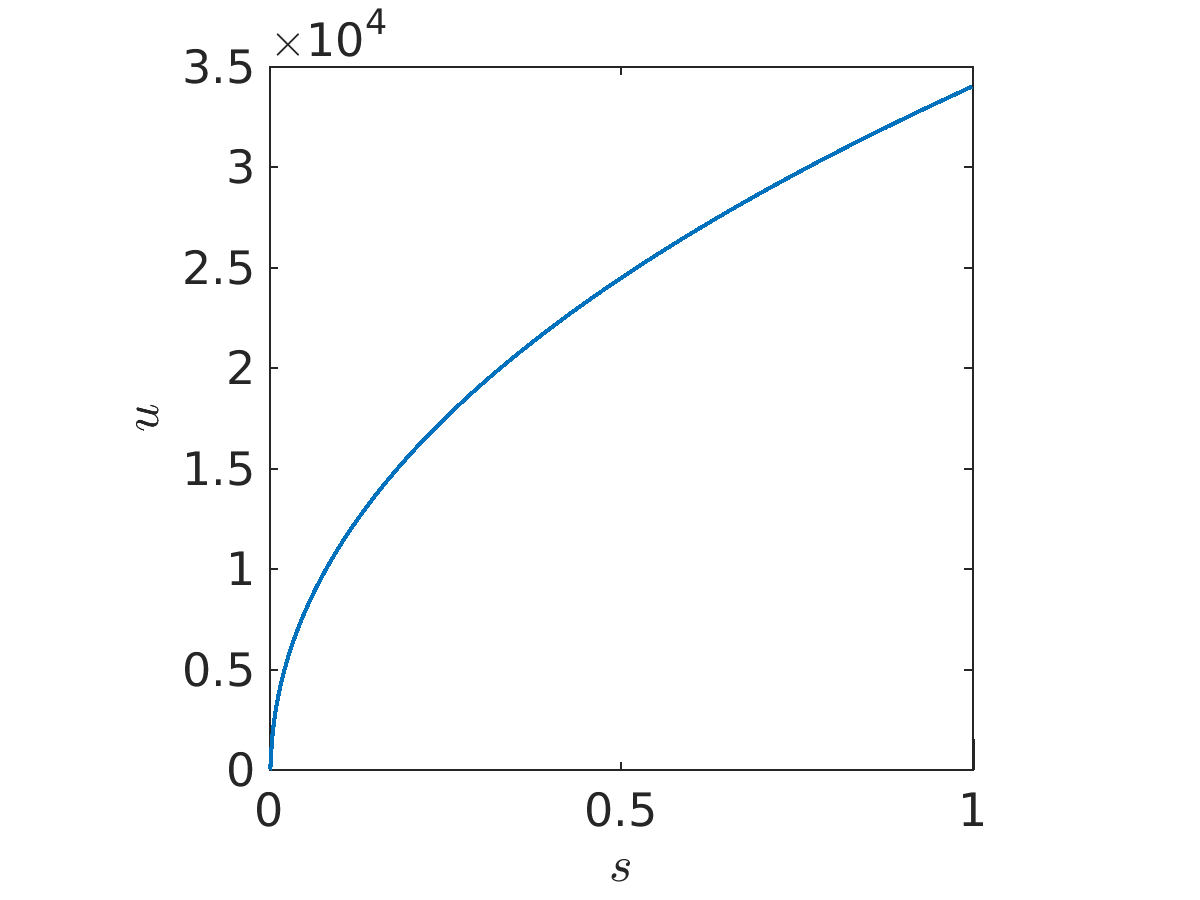}\\
\includegraphics[width=0.49\textwidth]{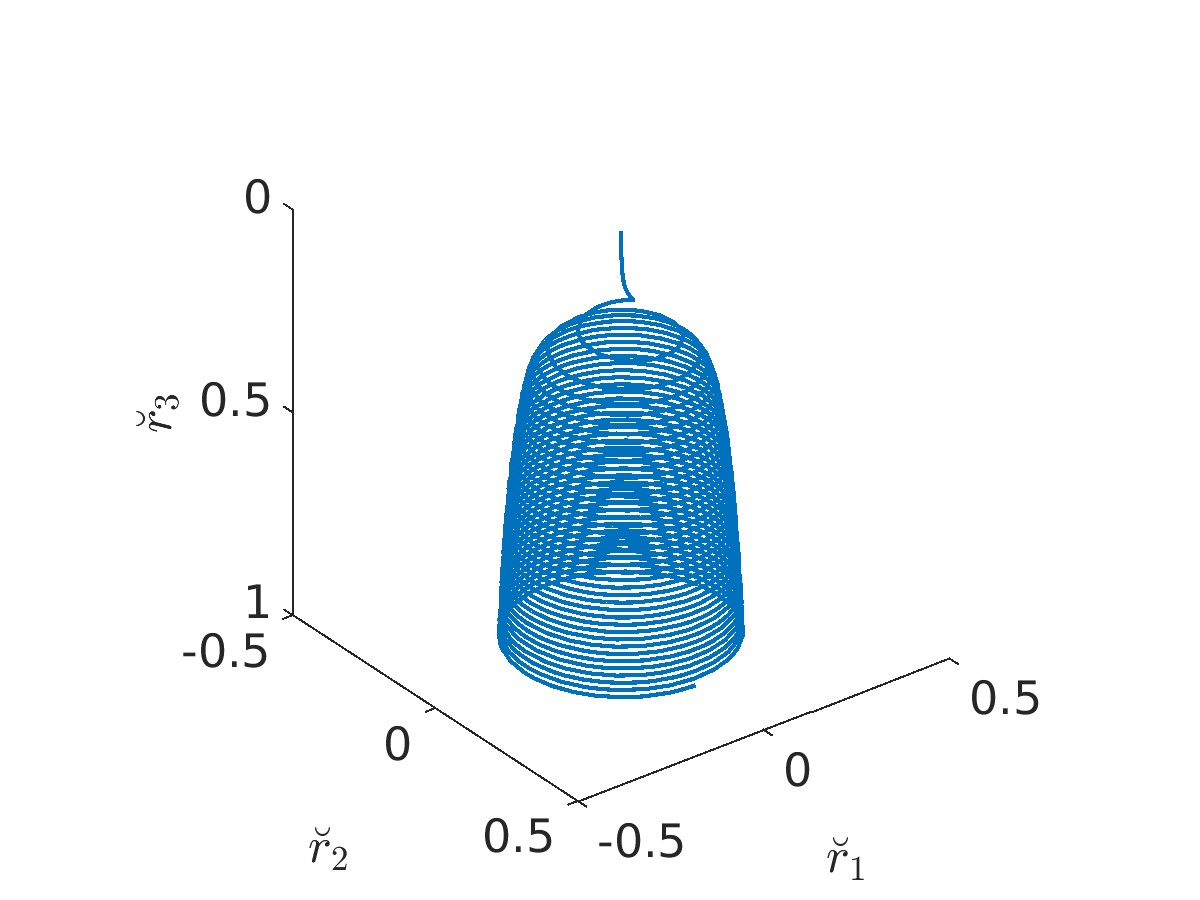}\hfill
\includegraphics[width=0.49\textwidth]{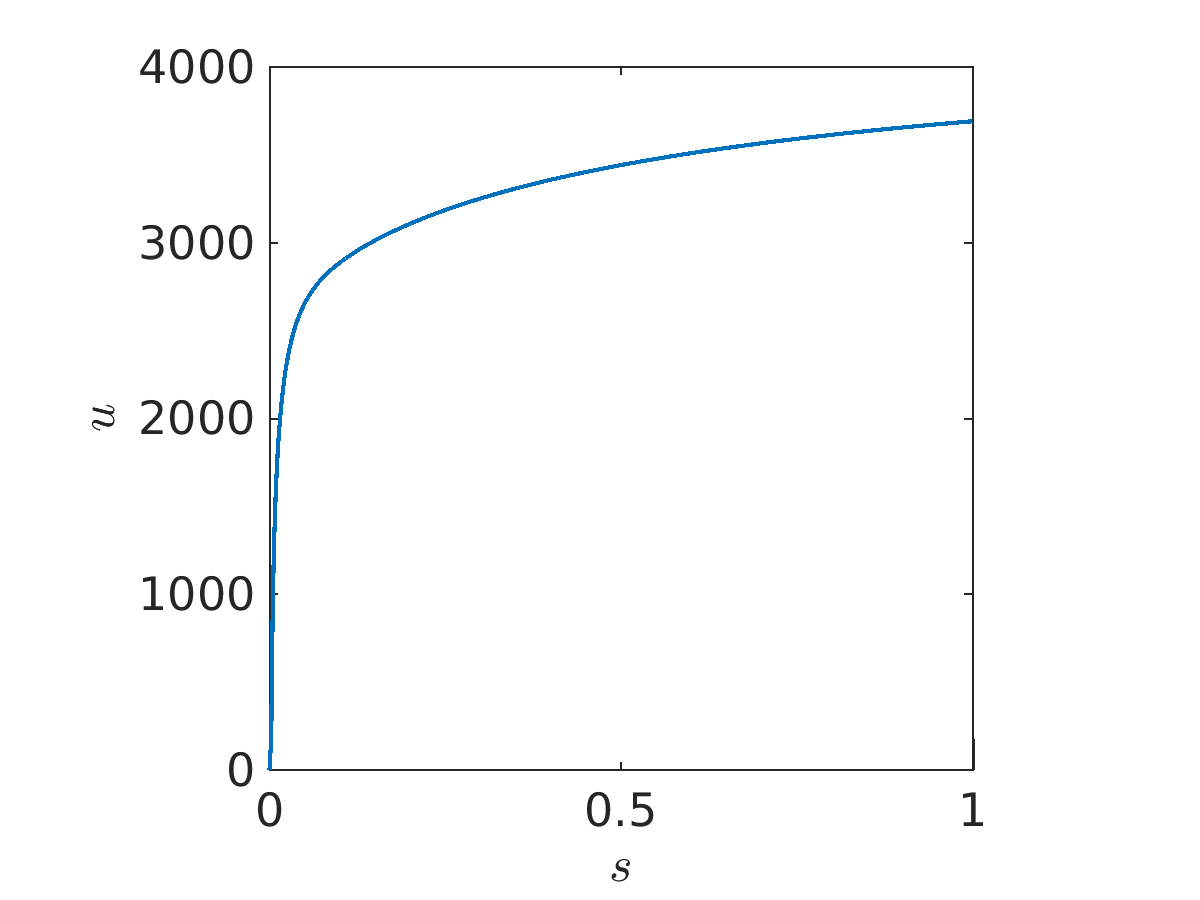}
\caption{\label{fig:highElong}\emph{Top:} Results for Example 1, $(\epsilon,\mathrm{M}) = (2\cdot 10^{-2},0)$. \emph{Bottom:} Results for Example 2, $(\epsilon,\mathrm{M}) = (7\cdot 10^{-2},0.1)$. \emph{Left:} Jet curve characterized by distinct whipping. \emph{Right:} Convective speed over arc length as indicator for the jet's elongation.}
\end{figure}

\begin{figure}[t]
\hspace*{-1.2cm}
\begin{minipage}[t]{0.32\textwidth}
\begin{center}
\includegraphics[width=1.3\textwidth]{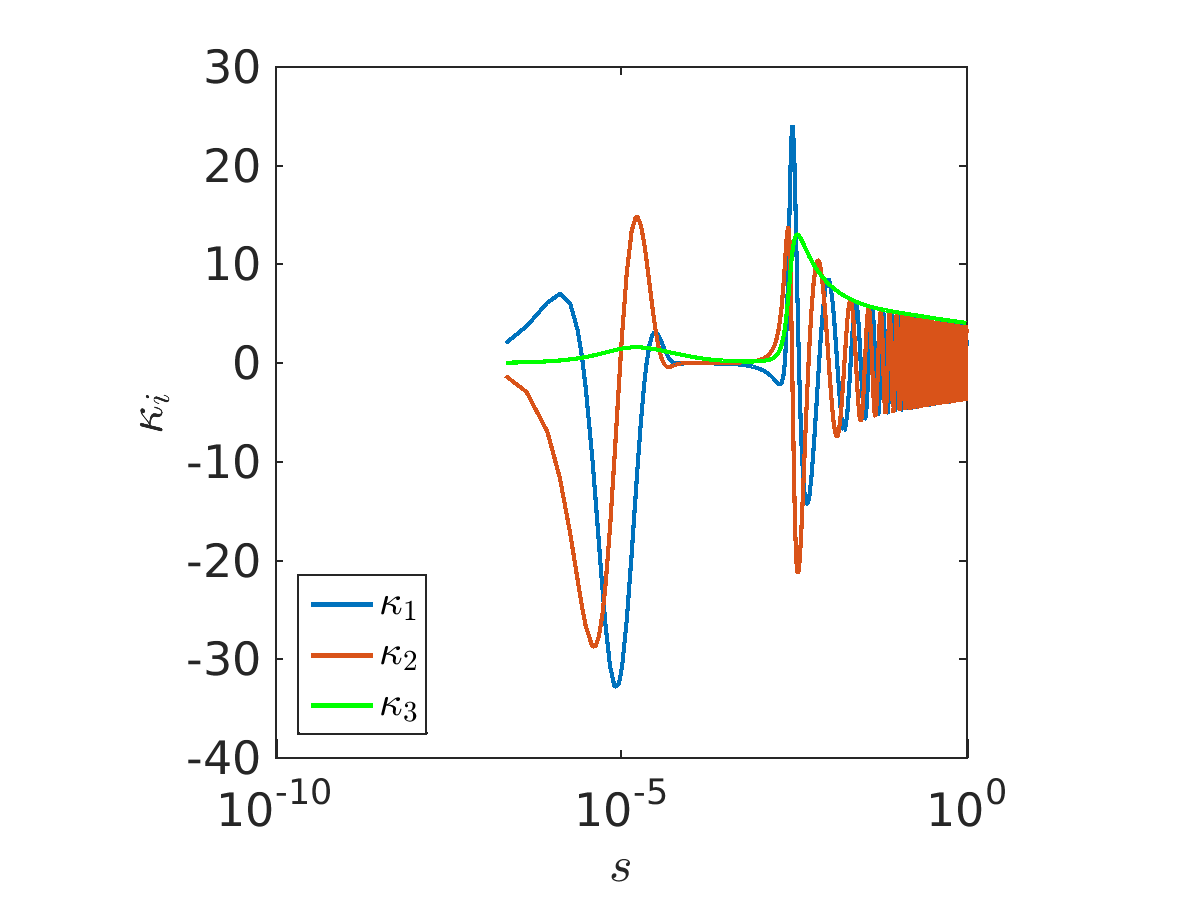}
\end{center}
\end{minipage}
\hspace*{0.2cm}
\begin{minipage}[t]{0.32\textwidth}
\begin{center}
\includegraphics[width=1.3\textwidth]{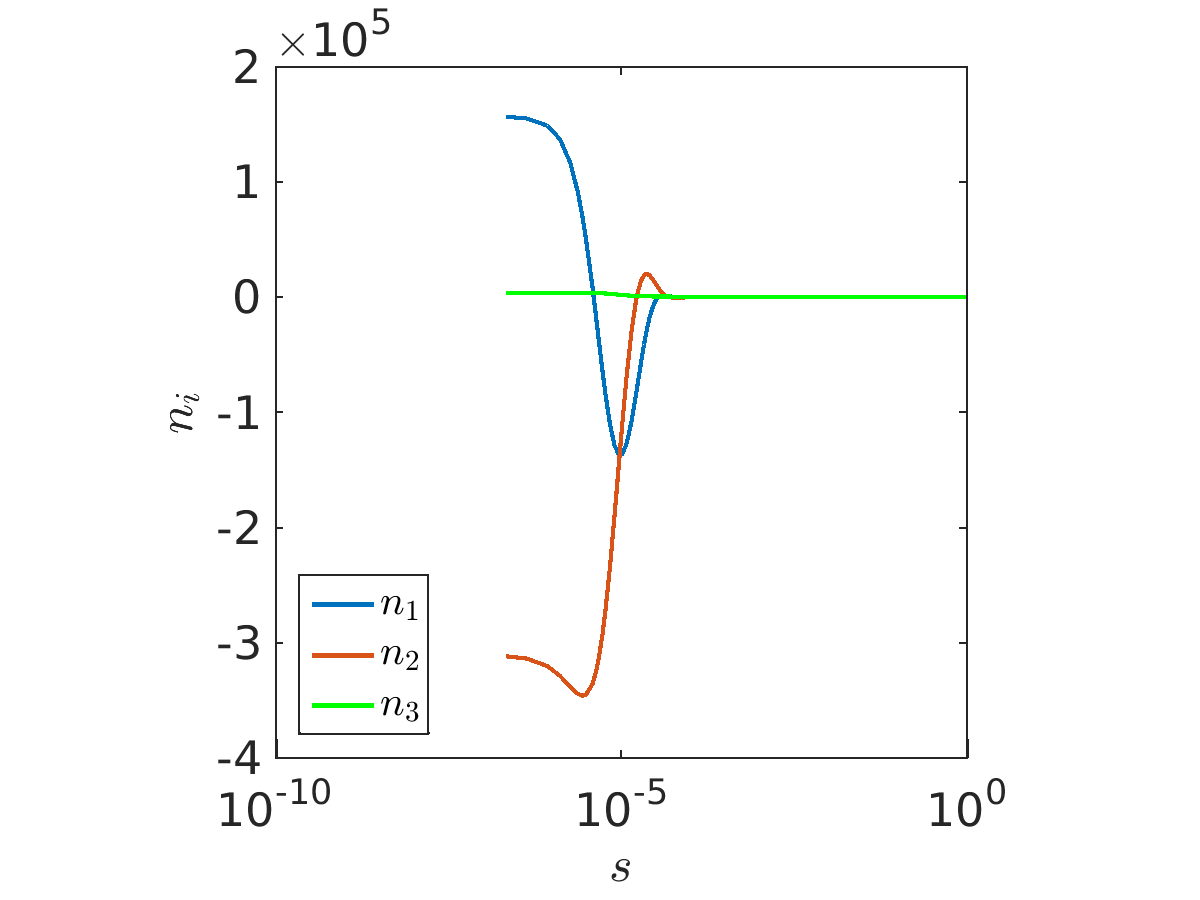}
\end{center}
\end{minipage}
\hspace*{0.2cm}
\begin{minipage}[t]{0.32\textwidth}
\begin{center}
\includegraphics[width=1.3\textwidth]{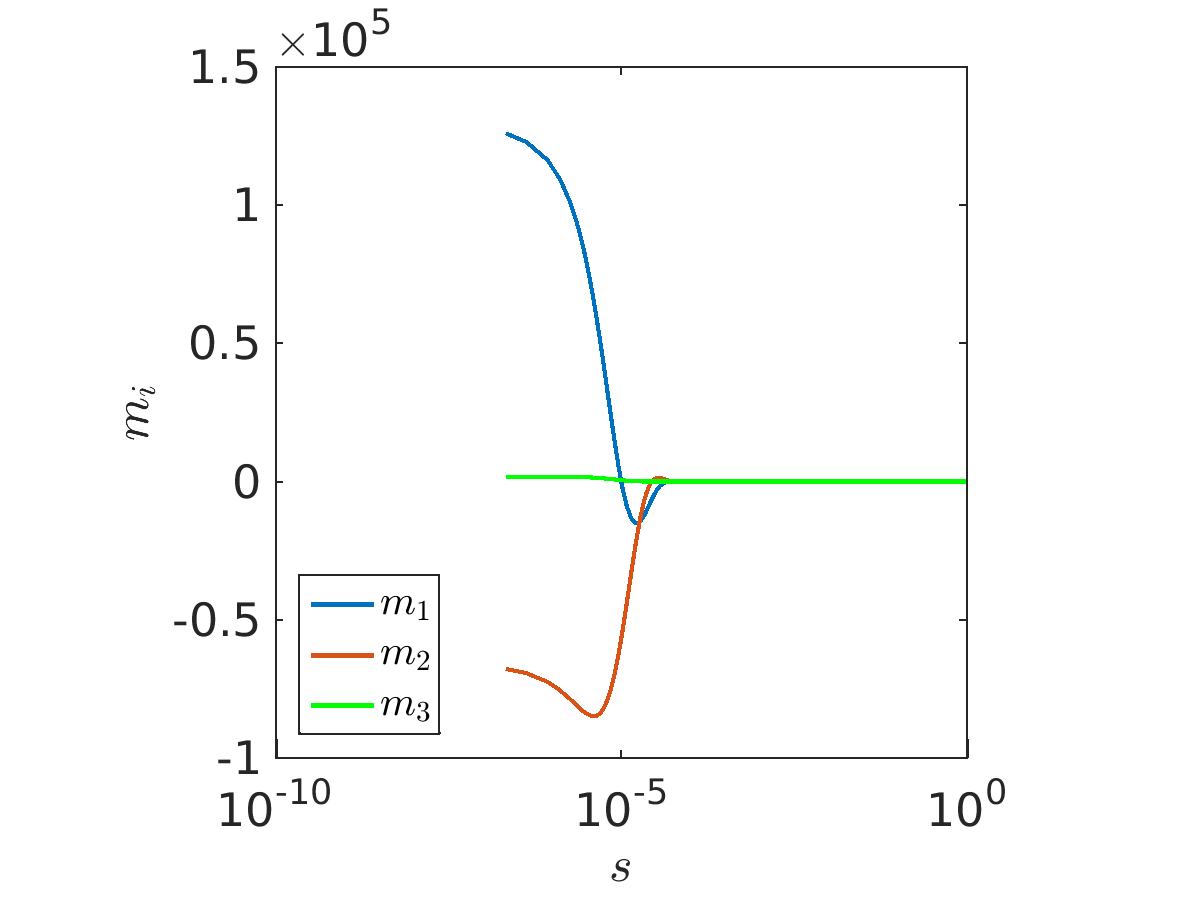}
\end{center}
\end{minipage}
\caption{\label{fig:kappaNMLayer}\emph{From left to right:} Boundary layer at the nozzle in the components of curvature $\mathsf{\kappa}$, contact force $\mathsf{n}$ and couple $\mathsf{m}$ for the parameters of Example~2.}
\end{figure}

Motivated from the parameter study in the previous subsection, we consider a setting with a highly developed whipping, namely $\mathrm{Re} = 3.5$, $\Gamma = 250$, $\Xi = 10^5$, $\Theta = 170$ and $\mathrm{Re}_\star = 1.5$. Concerning the remaining two dimensionless model parameters $\epsilon$ and $\mathrm{M}$ we point out: On the one hand our air model neglects any coupling between the jet and the surrounding air, that means we overestimate air drag forces in general. On the other hand as seen previously a high air drag-associated number $\mathrm{M}$ and a small slenderness ratio $\epsilon$ severally induce large numbers of turning points of the jet curve, which can easily require a high numerical resolution and thus lead to an unpractical number of discretization points in our collocation method. Thus we focus our considerations on two examples:
\begin{align*}
\text{Example 1:} \qquad &(\epsilon,\mathrm{M}) = (2\cdot 10^{-2},0)\\
\text{Example 2:} \qquad &(\epsilon,\mathrm{M}) = (7\cdot 10^{-2},0.1).
\end{align*}
In Example~1 any air drag-associated effects are neglected, see Fig.~\ref{fig:highElong} (top). 
The jet curve bends strongly and its envelope can nearly be described by a circular right cone. The dimensionless convective velocity $u$ indicating the jet's elongation describes a rapid growth directly at the nozzle with lower increase afterwards. At the jet end we reach approximately $u(1) = 3.4\cdot 10^4$ (compared to $u(0)=1$). In Example~2 the additional air drag causes a considerable increase of the number of turning points under decrease of the throwing range, see Fig. \ref{fig:highElong} (bottom). Due to the additional resistance caused by the surrounding air the convective speed remains in more moderate regions compared to the first example. In the components of curvature $\kappa$, contact force $\mathsf{n}$ and couple $\mathsf{m}$ a boundary layer arises directly at the nozzle (Figure~\ref{fig:kappaNMLayer}). This boundary layer is not a numerical artefact but an essential characteristic of our solution (see also App.~\ref{appendixB}).

\begin{figure}[t]
\includegraphics[width=0.6\textwidth]{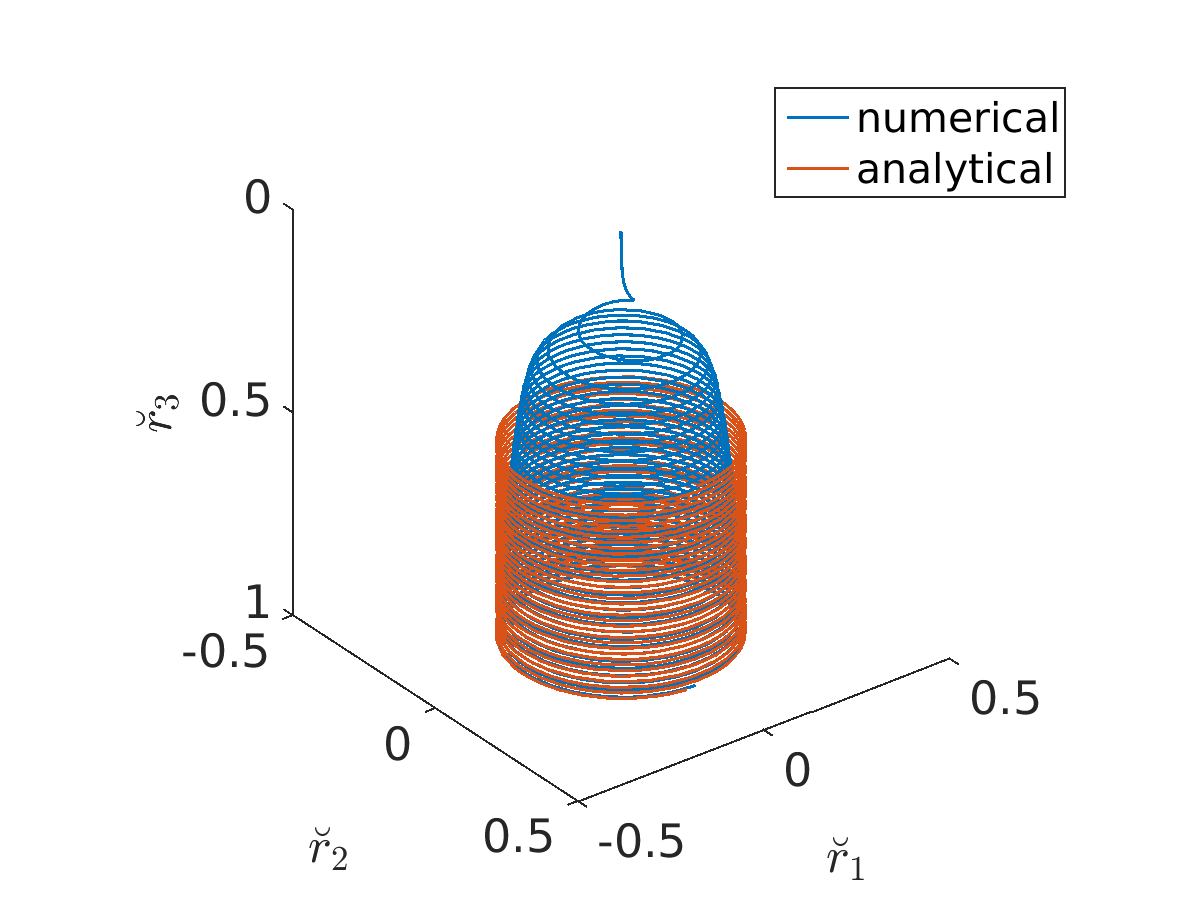}
\caption{\label{fig:analyticSol} Analytically described longtime behavior of the jet curve of Example 2. The numerical solution approaches the analytical description (\ref{eq:analyticSol}).}
\end{figure}

Moreover, the contact force $\mathsf{n}$ and couple $\mathsf{m}$ vanish beyond the boundary layer. This serves as motivation to consider the rod model in terms of vanishing contact forces and couples. Because the resulting solution is expected to describe the jet's 'longtime' behavior, we neglect all initial conditions and also leave the jet length $L$ and whipping frequency $\Omega$ arbitrary. Surprisingly, fixing $\mathsf{n} = \mathsf{m} = 0$ in our rod model (\ref{eq:model}) leads to an analytically solvable system. A solution thereof is
\begin{equation}
\begin{aligned}\label{eq:analyticSol}
& \breve{r}_1(s) = -\frac{2a_0a_1L}{b_0+b_1}\sin((b_0+b_1)s+c_0+c_1),\qquad
\breve{r}_2(s) = -\frac{2a_0a_1L}{b_0+b_1}\cos((b_0+b_1)s+c_0+c_1),\\
&\breve{r}_3(s) = L(a_0^2-a_1^2)s+d_3,\\
&q_0(s) = a_0\cos(b_0s+c_0),\qquad q_1(s) = a_1\sin(b_1s+c_1),\qquad
q_2(s) = a_1\cos(b_1s+c_1),\\
&q_3(s) = a_0\sin(b_0s+c_0),\\
&\kappa_1(s) = k_1\cos\bigg(\frac{\Omega L}{z}s+k_2\bigg),\quad\kappa_2(s) = -k_1\sin\bigg(\frac{\Omega L}{z}s+k_2\bigg),\qquad \kappa_3(s) = k_3,\\
&u(s) = z,
\end{aligned}
\end{equation}
with 13 parameters $\mathsf{x} = (a_0,a_1,b_0,b_1,c_0,c_1,d_3,k_1,k_2,k_3,z,\Omega,L) \in\mathbb{R}^{13}$ being implicitly given by the highly nonlinear system of equations
\begin{equation}
\begin{aligned}\label{eq:paramSys}
\frac{L}{2}\bigg(k_1a_1+k_3a_0\bigg) = -a_0b_0, \qquad\frac{L}{2}\bigg(-k_1a_0+k_3a_1\bigg) &= a_1b_1,\\
b_1-b_0 = \frac{\Omega L}{z},\qquad c_1-c_0 = k_2,\qquad a_0^2+a_1^2 &= 1,\\
\frac{\mathrm{Re}}{4z}\bigg(4a_0a_1\Omega k_3-4a_0a_1\frac{\Omega^2}{z}+2a_0a_1\frac{\Omega^2}{z}(a_0^2-a_1^2)+k_1 z k_3-k_1\Omega\bigg) &= 0,\\
\mathrm{Re} z+\frac{4\mathrm{Re}\Omega a_0a_1}{k_1}-\frac{2\Gamma}{\sqrt{z}}+\frac{\Xi\Theta}{z^2}\log\bigg(\frac{2\sqrt{z}}{\epsilon}\bigg)-\mathrm{Re}\frac{2a_0a_1\Omega^2L}{k_1z(b_0+b_1)} &= 0,\\
-\frac{4\Xi}{z}+\frac{r_\nu(w_\nu)}{w_\nu}\frac{2\mathrm{Re} \mathrm{M}\mathrm{Re}_\star^2|a_0a_1|L^2\Omega^2(1-4a_0^2a_1^2)}{\sqrt{z}(b_0+b_1)^2} &= 0,\\
-\frac{4\Xi}{z}(a_0^2-a_1^2)-\frac{r_\tau(w_\nu)}{w_\nu}\frac{2\mathrm{Re} \mathrm{M}\mathrm{Re}_\star^2|a_0a_1|L\Omega}{\sqrt{z}|b_0+b_1|}(a_0^2-a_1^2)\bigg(\frac{4a_0^2a_1^2\Omega L}{b_0+b_1}-z\bigg) &= 0,
\end{aligned}
\end{equation}
with air drag associated normal velocity components
\begin{align*}
w_\nu = \frac{2|\mathrm{Re}_\star a_0a_1L\Omega|}{\sqrt{z}|b_0+b_1|}|a_0^2-a_1^2|.
\end{align*}
With the particular choice
\begin{align*}
\mathsf{x} = (0.71,-0.7042,-9.1,235.2,0,0,0.507,3.77,0,4,3750,1.427\cdot 10^4,60.1)
\end{align*}
the system (\ref{eq:paramSys}) is approximately fulfilled. The corresponding solution (\ref{eq:analyticSol}) forms a helical jet curve that is approached by the numerical computed jet curve of Example 2, see Fig. \ref{fig:analyticSol}. So in the regime of distinct whipping we are able to give an analytical description of the jet's 'longtime' behavior.

Summing up, despite the simplicity of the employed electric force model where the current is considered as a constant parameter and conduction is neglected, the numerical results are convincing and characterize qualitatively well the known whipping effects.  However, for quantitative comparisons with experiments we lack respective measurements. Experimental studies in literature investigated intensively the whipping characteristics in dependence of various parameter settings,  but the underlying potential-current relations of the used set-ups are mostly not quantitatively documented. An exception is \cite{hohman:p:2001a, shin:p:2001} where experiments with a solvent of polyethylene oxide in water were performed. But due to the high conductivity of the solvent, we can not benefit from these results. A combined experimental and numerical study is left to future research.

\section{Conclusion}

In this paper we showed that the whipping instability observed in experiments with electrified visco-capillary jets can be directly approached with the stable stationary solutions of a Cosserat rod model. Our proposed continuation-collocation method for the respective boundary value problems assures efficient and accurate simulations including the automatic navigation through high-dimensional parameter spaces. We qualitatively verified the jet characteristics (whipping frequency, elongation, throwing range) to changes of the model parameters and showed the computability of parameter settings involving drastic jet thinning. Even analytical solutions for the 'longtime' behavior of the whipping could be derived. In view of quantitative predictions a combined experimental and numerical study is aimed at in future.

\appendix
\renewcommand*{\thesection}{\Alph{section}}
\renewcommand\thefigure{\thesection.\arabic{figure}}
\renewcommand\theequation{\thesection.\arabic{equation}}
\setcounter{section}{0}
\setcounter{figure}{0}
\setcounter{equation}{0}

\section{Global Linear Stability Analysis} \label{appendixA}

The whipping instability observed in experiments was investigated by means of classical perturbation theory in e.g.\ \cite{hohman:p:2001,reneker:p:2000, yarin:p:2001, li:p:2009}. In this context, we point out that the unstable solutions of our transient rod model \eqref{eq:rod} are directly related to the whipping instability as we can conclude from the following global stability analysis. 

Proceeding from the time-dependent rod system (\ref{eq:rod}) in which the geometric model, the material laws and the capillary force are inserted we replace the kinematic equations for the time derivatives of curve and triad by the compatibility conditions \cite{arne:p:2010}, yielding
\begin{equation}\label{eq:rod_time}
\begin{aligned}
\partial_s \mathbf{r} = \mathbf{d_3},  \qquad\qquad &
\partial_s \mathbf{d_i} = \boldsymbol{\kappa} \times \mathbf{d_i},\\
\partial_s (\mathbf{v} - u\mathbf{d_3}) = (\boldsymbol{\omega}-u\boldsymbol{\kappa}) \times \mathbf{d_3}, \qquad\qquad &
\partial_t \boldsymbol{\kappa} + \partial_s(u\boldsymbol{\kappa} - \boldsymbol{\omega}) = \boldsymbol{\omega}\times\boldsymbol{\kappa},\\
\partial_t (a^2) + \partial_s (u a^2) &= 0, \\
\partial_t (\rho \pi a^2 \mathbf{v}) + \partial_s (\rho\pi u a^2\mathbf{v} - \mathbf{n} - \pi\gamma a\mathbf{d_3}) &= \mathbf{f}_{el} + \mathbf{f}_{air},\\
 \partial_t (\frac{\rho\pi}{4}a^4\mathbf{P}_2 \cdot \boldsymbol{\omega}) +  \partial_s (\frac{\rho\pi}{4}  ua^4 \mathbf{P}_2 \cdot \boldsymbol{\omega} - \mathbf{m}) &= \mathbf{d_3}\times \mathbf{n},\\
\partial_s u = \frac{1}{3\mu \pi a^2}\mathbf{n}\cdot\mathbf{d_3}, \qquad\qquad & \partial_s\boldsymbol{\omega} = \frac{4}{3\mu\pi a^4}\mathbf{P}_{3/2}\cdot\mathbf{m}
\end{aligned}
\end{equation}
with the electric and aerodynamic forces
\begin{align*}
\mathbf{f}_{el} = 2\pi a\sigma \bigg(\mathbf{E}-\frac{a\sigma}{2\varepsilon_p}\log\bigg(\frac{H}{a}\bigg)\boldsymbol{\kappa}\times\mathbf{d_3}\bigg),\qquad\qquad
\mathbf{f}_{air} = \frac{\mu_\star^2}{2a\rho_\star}\mathbf{F}\bigg(\boldsymbol{\mathbf{d_3}},-\frac{2a\rho_\star}{\mu_\star}\mathbf{v}\bigg).
\end{align*}
We supplement the system with the spinning-associated boundary conditions, i.e., inflow at nozzle ($s=0$) and stress-free at the jet end ($s=L$), keeping in mind that the jet length $L$ is an unknown and requires an additional condition,
\begin{align*}
\mathbf{r}(0,t) &= \mathbf{0},&\quad \mathbf{d_i}(0,t) &= \mathbf{a_i^\circ} ,&\quad \mathbf{v}(0,t) &= U\mathbf{d_3} ,&\quad \boldsymbol{\kappa}(0,t) &= \mathbf{0},&\quad a(0,t) &= D/2,\\
\quad \mathbf{n}(L,t) &= \mathbf{0},&\quad \mathbf{m}(L,t) &= \mathbf{0},& \quad u(0,t) &= U,&\quad \boldsymbol{\omega}(0,t) &= \mathbf{0},&\quad \mathbf{r}(L,t)\cdot \mathbf{a_3^\circ}&=H.
\end{align*}
In particular we consider here a stationary outer basis $\{\mathbf{a_1^\circ},\mathbf{a_2^\circ},\mathbf{a_3^\circ}\}$ with predominant spinning direction $\mathbf{a_3^\circ} = \mathbf{a_3}$. To preserve the time-dependencies we formulate \eqref{eq:rod_time} with respect to the director basis and the stationary outer basis using the coordinate terminology $\mathbf{y} = \sum_{i=1}^3\hat{y}_i\mathbf{d_i} = \sum_{i=1}^3 \breve{y}_i^\circ\mathbf{a_i}^\circ$
with $\hat{\mathsf{y}} = (\hat{y}_1,\hat{y}_2,\hat{y}_3)\in\mathbb{R}^3$ and $\breve{\mathsf{y}}^\circ = (\breve{y}_1^\circ,\breve{y}_2^\circ,\breve{y}_3^\circ)\in\mathbb{R}^3$. The corresponding tensor-valued rotation is denoted by $\hat{\mathbf{R}}=\mathbf{a_i}^\circ\otimes\mathbf{d_i}$ with associated matrix $\hat{\mathsf{R}}$. To the dimensionless quantities $\mathsf{\tilde{y}}(\tilde{s},\tilde{t}) = \mathsf{y}(\bar{s}\tilde{s},\bar{t}\tilde{t})/\bar{y}$ of Sec.~\ref{sec:2} we introduce the additional reference values $\bar{v} = U$, $\bar{a} = D$, $\bar{\omega} = U/H$ and $\bar{t} = L/U$. Suppressing the label $\tilde{~}$, the respective dimensionless system reads
\begin{equation*}
\begin{aligned}
L^{-1}\partial_s \breve{\mathsf{r}}^\circ &= \hat{\mathsf{R}}^T\cdot\mathsf{e_3},\\
L^{-1}\partial_s \hat{\mathsf{R}} &= -\hat{\mathsf{\kappa}}\times\hat{\mathsf{R}},\\
L^{-1}\partial_s(\hat{\mathsf{v}}-u\mathsf{e_3}) &= \hat{\mathsf{\omega}}\times\mathsf{e_3} + \hat{\mathsf{v}}\times\hat{\mathsf{\kappa}},\\
L^{-1}(\partial_t\hat{\mathsf{\kappa}} + \partial_s(u\hat{\mathsf{\kappa}}-\hat{\mathsf{\omega}})) &= \hat{\mathsf{\kappa}}\times\hat{\mathsf{\omega}},\\
\partial_t(a^2)+\partial_s(u a^2) &=0,\\
L^{-1}\bigg(\partial_t(a^2\hat{\mathsf{v}}) + \partial_s\bigg(u a^2\hat{\mathsf{v}} - \frac{1}{4\mathrm{Re}}\hat{\mathsf{n}} - \frac{\Gamma}{\mathrm{Re}} a\mathsf{e_3}\bigg)\bigg) &= a^2\hat{\mathsf{v}}\times\hat{\mathsf{\omega}} + \frac{1}{4\mathrm{Re}}\hat{\mathsf{\kappa}}\times\hat{\mathsf{n}} + \frac{\Gamma}{\mathrm{Re}}a \hat{\mathsf{\kappa}}\times\mathsf{e_3} + \hat{\mathsf{f}}_{el} + \hat{\mathsf{f}}_{air},\\
L^{-1}\bigg(\partial_t(a^4\mathsf{P}_2\cdot\hat{\mathsf{\omega}}) + \partial_s\bigg(u a^4\mathsf{P}_2\cdot\hat{\mathsf{\omega}}-\frac{1}{4\mathrm{Re}}\hat{\mathsf{m}}\bigg)\bigg) &= a^4(\mathsf{P}_2\cdot\hat{\mathsf{\omega}})\times\hat{\mathsf{\omega}} + \frac{1}{4\mathrm{Re}}\hat{\mathsf{\kappa}}\times\hat{\mathsf{m}} + \frac{1}{\varepsilon^2\mathrm{Re}}\mathsf{e_3}\times\hat{\mathsf{n}},\\
L^{-1}\partial_s u &= \frac{1}{12}\frac{\hat{n}_3}{a^2},\\
L^{-1}\partial_s \hat{\mathsf{\omega}} &= \hat{\mathsf{\omega}}\times\hat{\mathsf{\kappa}} + \frac{1}{12}\frac{1}{a^4}\mathsf{P}_{3/2}\cdot\hat{\mathsf{m}},
\end{aligned}
\end{equation*}
with the outer forces
\begin{align*}
\hat{\mathsf{f}}_{el} = \frac{\Xi}{\mathrm{Re}}\bigg(\frac{1}{u}\hat{\mathsf{R}}\cdot\mathsf{e_3}-\frac{\Theta}{4}\frac{1}{u^2}\log\bigg(\frac{1}{\epsilon a}\bigg)\hat{\mathsf{\kappa}}\times\mathsf{e_3}\bigg),\qquad\qquad
\hat{\mathsf{f}}_{air} = \frac{\mathrm{M}}{8}\frac{1}{a}\hat{\mathsf{F}}(\mathsf{e_3},-2\mathrm{Re}_\star a\hat{\mathsf{v}}),
\end{align*}
and boundary conditions
\begin{align*}
\breve{\mathsf{r}}^\circ(0,t) &= 0,&\quad \hat{\mathsf{R}}(0,t) &= \mathsf{P}_1,&\quad \hat{\mathsf{v}}(0,t) &= \mathsf{e_3} ,&\quad \hat{\mathsf{\kappa}}(0,t) &= 0,&\quad a(0,t) &= 1/2,&\\
\quad \hat{\mathsf{n}}(1,t) &= 0,&\quad \hat{\mathsf{m}}(1,t) &= 0,& \quad u(0,t) &= 1,&\quad \hat{\mathsf{\omega}}(0,t) &= 0, &\quad \breve{r}_3^\circ(1,t) &= 1.
\end{align*}

Obviously, the boundary value problem has the general conservation form
\begin{align}\label{eq:bvp}
\partial_t {\mathsf{h}}(\hat{\mathsf{y}}(s,t)) + \partial_s {\mathsf{j}}(\hat{\mathsf{y}}(s,t)) = {\mathsf{k}}(\hat{\mathsf{y}}(s,t)),\qquad\qquad
{\mathsf{g}}(\hat{\mathsf{y}}(0,t),\hat{\mathsf{y}}(1,t)) = 0.
\end{align}
For the temporal stability analysis we assume the solution form
\begin{align*}
\hat{\mathsf{y}}(s,t) = \hat{\mathsf{y}}_\varepsilon(s,t) = \hat{\mathsf{y}}_0(s)+\varepsilon e^{\lambda t}\hat{\mathsf{y}}_1(s)
\end{align*}
with $\lambda\in\mathbb{C}$ and $0 < \varepsilon \ll 1$. By means of this ansatz and Taylor expansion around $\hat{\mathsf{y}}_0$, \eqref{eq:bvp} can be split into two systems, i.e.,
\begin{align}\label{eq:y0}
\frac{\mathrm{d}}{\mathrm{d}s}\mathsf{j}(\hat{\mathsf{y}}_0(s)) = \mathsf{k}(\hat{\mathsf{y}}_0(s)), \qquad \mathsf{g}(\hat{\mathsf{y}}_0(0),\hat{\mathsf{y}}_0(1)) = 0,
\end{align}
for the stationary solution $\hat{\mathsf{y}}_0$ as well as
\begin{align}\label{eq:y1}
\bigg(\mathsf{K}(s) - \frac{\mathrm{d}}{\mathrm{d}s} \mathsf{J}(s) - \mathsf{J}(s) \frac{\mathrm{d}}{\mathrm{d}s} \bigg)\cdot\hat{\mathsf{y}}_1(s) = \lambda \mathsf{H}(s)\cdot\hat{\mathsf{y}}_1(s),\qquad
\mathsf{A}\cdot\hat{\mathsf{y}}_1(0) + \mathsf{B}\cdot\hat{\mathsf{y}}_1(1) = 0,
\end{align}
for the transient correction $\hat{\mathsf{y}}_1$ and $\lambda$, where $\mathsf{H}(s) = \partial_{\hat{\mathsf{y}}} {\mathsf{h}}(\hat{\mathsf{y}}_0(s))$, $\mathsf{J}(s) = \partial_{\hat{\mathsf{y}}} {\mathsf{j}}(\hat{\mathsf{y}}_0(s))$,
$\mathsf{K}(s) = \partial_{\hat{\mathsf{y}}} {\mathsf{k}}(\hat{\mathsf{y}}_0(s))$,
$\mathsf{A} = \partial_1 {\mathsf{g}}(\hat{\mathsf{y}}_0(0),\hat{\mathsf{y}}_0(1))$ and
$\mathsf{B} = \partial_2 {\mathsf{g}}(\hat{\mathsf{y}}_0(0),\hat{\mathsf{y}}_0(1))$. Thereby, the sign of the real part of $\lambda$ decides about stability or instability of $\hat{\mathsf{y}}_0$ for $t\rightarrow\infty$.

A solution of (\ref{eq:y0}) is obviously a stationary jet that forms a straight line from the nozzle towards the collector, i.e.,
\begin{align*}
\breve{\mathsf{r}}^\circ_0(s) &= s\mathsf{e_3},&\quad \hat{\mathsf{R}}_0(s) &= \mathsf{P}_1,&\quad \hat{\mathsf{v}}_0(s) &= u_0(s)\mathsf{e_3},&\quad \hat{\mathsf{\kappa}}_0(s) &= 0 ,&\quad
a_0(s) &= (2\sqrt{u_0(s)})^{-1},&\\
\hat{\mathsf{n}}_0(s) &= \hat{n}_{0,3}(s)\mathsf{e_3},&\quad \hat{\mathsf{m}}_0(s) &= 0,&\quad \hat{\mathsf{\omega}}_0(s) &= 0,&\quad L=1&
\end{align*}
with $u_0$, $\hat{n}_{0,3}$ being prescribed by the boundary value problem
\begin{align*}
\partial_s u_0 &= \frac{1}{3} u_0 \hat{n}_{0,3},&\quad\partial_s\hat{n}_{0,3} &= \frac{\mathrm{Re}}{3} u_0 \hat{n}_{0,3} + \frac{\Gamma}{3}\frac{\hat{n}_{0,3}}{\sqrt{u_0}} - 4\Xi\frac{1}{u_0} - \hat{\mathsf{f}}_{air}\cdot\mathsf{{e_3}},&\\
u_0(0) &= 1,&\quad \hat{n}_{0,3}(1) &= 0.&
\end{align*}
We determine $u_0$ and $\hat{\mathsf{n}}_{0,3}$ with the continuation-collocation method of Sec.~\ref{sec:3} and initial values $u_0 = 1$, $\hat{\mathsf{n}}_{0,3} = 0$. In order to solve (\ref{eq:y1}) we consider an equidistant grid of mesh size $h$ with grid points $s_i$, $i\in\{0,...,N\}$. Integrating the differential equation over the intervals $[s_{i-1},s_i]$ we apply the fundamental theorem of calculus and the trapezoidal quadrature rule respectively
\begin{align*}
0 &= \int\limits_{s_{i-1}}^{s_i}\lambda \mathsf{H}(s)\cdot\hat{\mathsf{y}}_1(s) + \partial_s \big(\mathsf{J}(s)\cdot\hat{\mathsf{y}}_1(s)\big)-\mathsf{K}(s)\cdot\hat{\mathsf{y}}_1(s)~\mathrm{d}s\\
&\approx\lambda\frac{h}{2}\big(\mathsf{H}_i\cdot\hat{\mathsf{y}}_{1,i}+\mathsf{H}_{i-1}\cdot
\hat{\mathsf{y}}_{1,i-1}\big) + \mathsf{J}_i\cdot\hat{\mathsf{y}}_{1,i} - \mathsf{J}_{i-1}\cdot\hat{\mathsf{y}}_{1,i-1}-\frac{h}{2}\big(\mathsf{K}_i\cdot\hat{\mathsf{y}}_{1,i}+\mathsf{K}_{i-1}
\cdot\hat{\mathsf{y}}_{1,i-1}\big)
\end{align*}
with $\hat{\mathsf{y}}_{1,i}=\hat{\mathsf{y}}_1(s_i)$, $\mathsf{H}_i=\mathsf{H}(s_i)$, $\mathsf{J}_i=\mathsf{J}(s_i)$, and $\mathsf{K}_i=\mathsf{K}(s_i)$.
After inclusion of the boundary condition we obtain a generalized eigenvalue problem \begin{align*}
{\mathsf{W}}\cdot \hat{\mathsf{Y}}=\lambda {\mathsf{Z}}\cdot \hat{\mathsf{Y}}, \qquad \mathsf{W},\mathsf{Z}\in\mathbb{R}^{N_v(N+1)}\otimes\mathbb{R}^{N_v(N+1)}
\end{align*} 
with
\begin{align*}
\mathsf{W}&=\begin{pmatrix}
\mathsf{A} & 0 & \cdots & & 0 & \mathsf{B}\\
\mathsf{J}_0+\frac{h}{2}\mathsf{K}_0 & -\mathsf{J}_1+\frac{h}{2}\mathsf{K}_1 & 0 & \cdots & & 0\\
0 & \mathsf{J}_1+\frac{h}{2}\mathsf{K}_1 & -\mathsf{J}_2+\frac{h}{2}\mathsf{K}_2 & 0 & \cdots  & 0\\
\vdots & & \ddots & \ddots & & \vdots\\
\vdots & & & \ddots & \ddots & 0\\
0 & \cdots & & 0 & \mathsf{J}_{N-1}+\frac{h}{2}\mathsf{K}_{N-1} & -\mathsf{J}_N+\frac{h}{2}\mathsf{K}_N
\end{pmatrix}\\
&\\
\mathsf{Z}&=\begin{pmatrix}
0 & 0 & \cdots & & 0 & 0\\
\frac{h}{2}\mathsf{H}_0 & \frac{h}{2}\mathsf{H}_1 & 0 & \cdots & & 0\\
0 & \frac{h}{2}\mathsf{H}_1 & \frac{h}{2}\mathsf{H}_2 & 0 & \cdots  & 0\\
\vdots & & \ddots & \ddots & & \vdots\\
\vdots & & & \ddots & \ddots & 0\\
0 & \cdots & & 0 & \frac{h}{2}\mathsf{H}_{N-1} & \frac{h}{2}\mathsf{H}_N
\end{pmatrix}
\end{align*}
and $N_v$ the number of rod unknowns. We compute the solution using the MATLAB routine $\emph{eig.m}$. Since $Z$ is singular it is convenient to consider the inverse problem $\tilde{\lambda}\mathsf{W}\cdot\hat{\mathsf{Y}} = \mathsf{Z}\cdot\hat{\mathsf{Y}}$, $\tilde{\lambda} = 1/\lambda$ transforming infinite eigenvalues to zero. 

\begin{figure}[!t]
\includegraphics[width=0.49\textwidth]{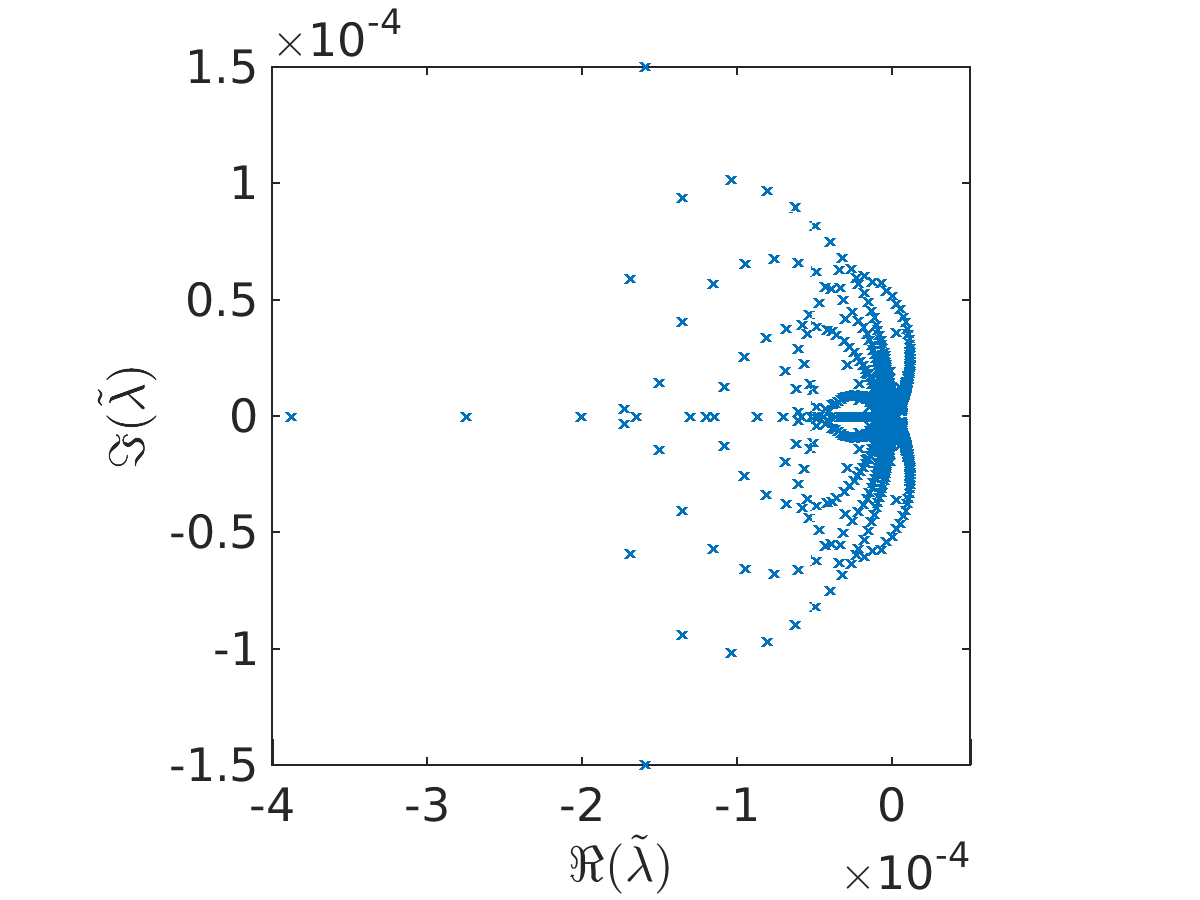}
\hfill
\includegraphics[width=0.49\textwidth]{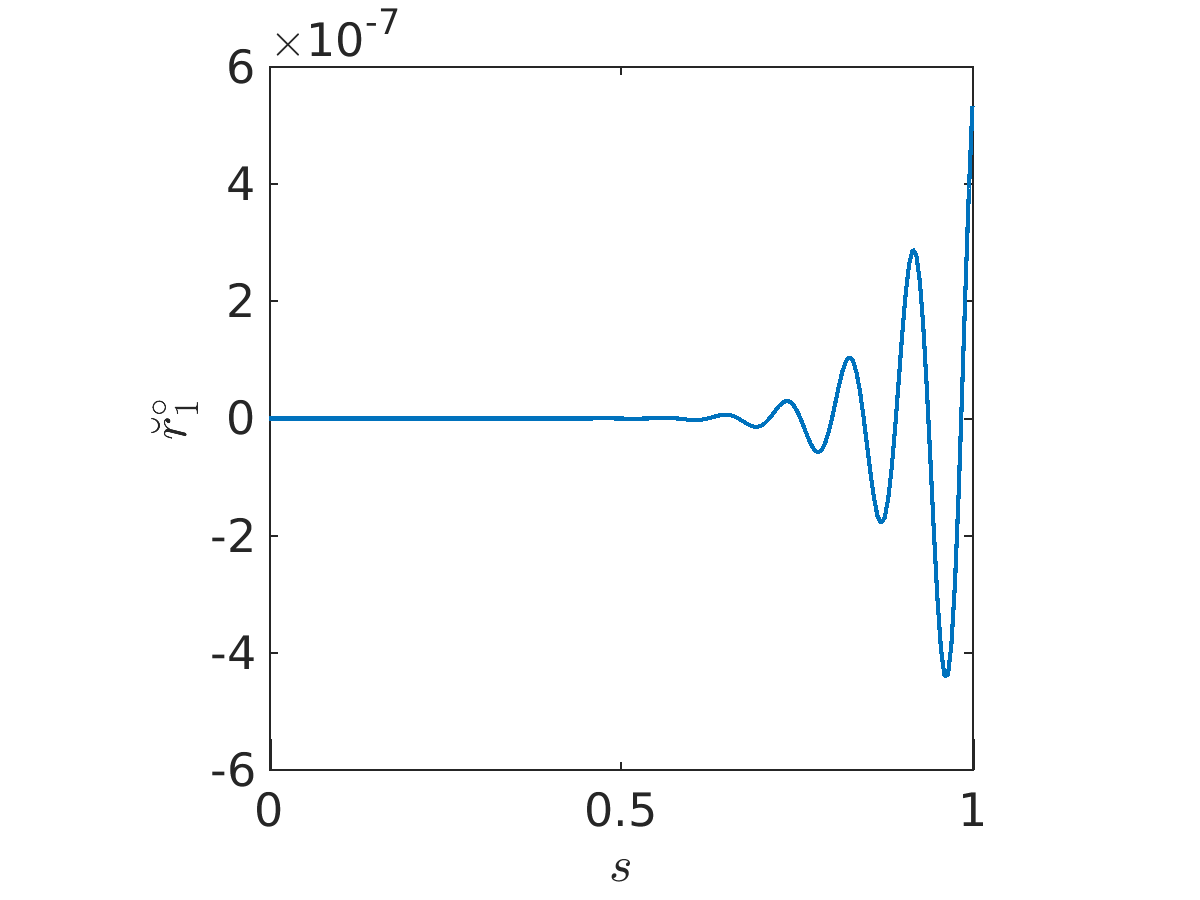}
\caption{\label{fig:stabAnalysis}Results of stability analysis for Example~2 and $N=200$. \emph{Left:} Spectrum of the inverse eigenvalues $\tilde{\lambda}$. \emph{Right:} Eigenfunction $\breve{r}_1^\circ$ related to the eigenvalue $\tilde{\lambda}$ with largest real part.}
\end{figure}

The observed whipping effect turns out to be a instability in the stated mathematical sense. Figure~{\ref{fig:stabAnalysis} (left) shows the eigenvalue spectrum to the model parameters of Example~2, belonging to a mesh with $N=200$. Obviously there exist (inverse) eigenvalues $\tilde{\lambda}$ with positive real part. The corresponding solution $\hat{\mathsf{y}}_\varepsilon$ moves away from the stationary solution $\hat{\mathsf{y}}_0$ for infinite times, which is hence unstable. 
The (discrete) eigenfunction $\breve{r}_1^\circ$ related to the eigenvalue $\tilde{\lambda}$ with largest real part reveals the formation of the characteristic whipping, see Fig.~\ref{fig:stabAnalysis} (right). We employ the transition from stable to unstable solutions in dependence on the electric force related parameters $\Xi$ and $\Theta$. For the computation of the phase diagram we choose discrete test parameters $\Xi_i = 1000i$, $i=0,...,100$ and determine the corresponding smallest $\Theta_i$, for which a solution of the eigenvalue problem with $N=200$ has an eigenvalue with positive real part. In Fig.~\ref{fig:stabDiagram} the lower region bounded by the data points and the horizontal axis characterizes the stable solutions, the upper region above the data points corresponds to the unstable solutions. As we have seen in Sec.~\ref{sec:4} an increasing outer electric field stabilizes the jet, hence for larger values of $\Xi$ the transition from stable to unstable solutions takes place for larger values of $\Theta$.

\begin{figure}[t]
\includegraphics[width=0.5\textwidth]{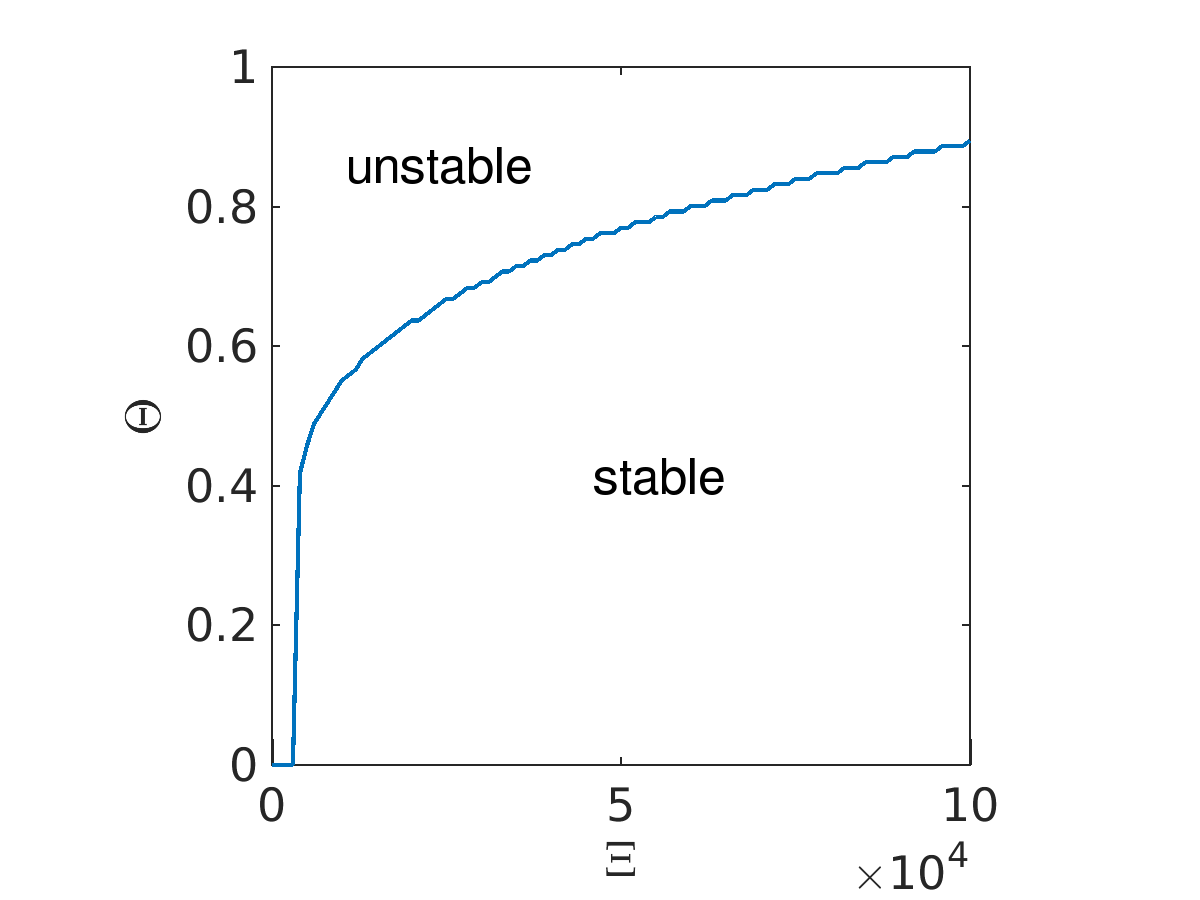}
\caption{\label{fig:stabDiagram}Stability diagram with respect to $(\Xi,\Theta)\in[0,10^5]\times\mathbb{R}^+_0$ for fixed parameters $(\mathrm{Re},\Gamma,\epsilon,\mathrm{M},\mathrm{Re}_\star) = (3.5,250,7\cdot 10^{-2},0.1,1.5)$ (cf.\ Example~2). The lower region indicates stable solutions, whereas the upper region characterizes unstable solutions.}
\end{figure}

\section{Boundary Layers} \label{appendixB}
\setcounter{figure}{0}
\setcounter{equation}{0}

For distinct whipping boundary layers arise in the variables $\kappa$, $\mathsf{n}$ and $\mathsf{m}$ at the nozzle, as seen in Fig.~\ref{fig:kappaNMLayer} for Example~2. To verify that the layers are model-based and that our numerical scheme resolves the effect correctly, we re-compute the results in material (Lagrangian) description.

For the transformation of our model \eqref{eq:model} from Eulerian to Lagrangian description, we apply the same concepts as in \cite{arne:p:2010}: We introduce a general bijective mapping $\Phi(\cdot,t): (\sigma_a(t),\sigma_b(t))\rightarrow(s_a(t),s_b(t))$, $\sigma\mapsto\Phi(\sigma,t)$ that satisfies the relations
\begin{align*}
\partial_t \Phi(\sigma,t) = u(\Phi(\sigma,t)), \qquad\Phi(\sigma,t_{in}(\sigma)) = 0
\end{align*}
where $t_{in}(\sigma)$ prescribes the time of the material point $\sigma$ entering the steady flow domain $\Phi(\sigma,t)\in[0,L]$. Thus, $\Phi$ depends only on the run time $\zeta(\sigma)=t-t_{in}(\sigma)$ setting $\Phi(\sigma,t) = \hat{\Phi}(t-t_{in}(\sigma)) = \hat{\Phi}(\zeta(\sigma))$. To ensure that all physical meanings are preserved in the material description we use the concept of 
type-$n$-fields, $n\in\mathbb{Z}$, i.e.\ to an arbitrary field $f$ in Eulerian description the associated Lagrangian field $\tilde{f}$ is related according to 
\begin{align*}
j^n(\sigma,t)f(\Phi(\sigma,t)) = \tilde{f}(\sigma,t), \quad j(\sigma,t) = \partial_\sigma\Phi(\sigma,t).
\end{align*}
Thereby, our rod variables $\breve{\mathsf{r}}$, $\mathsf{R}$, $\mathsf{v}$, $\mathsf{\omega}$, $\mathsf{n}$, $\mathsf{m}$, $J$, $A$ are treated as type-$0$-fields and $\mathsf{\kappa}$, $\mathsf{f}$, $j$ type-$1$-fields.
The exclusive dependence on the run time $\zeta$ is also valid for an arbitrary Lagrangian field, $\tilde{f}(\sigma,t) = \hat{f}(t-t_{in}(\sigma)) = \hat{f}(\zeta(\sigma))$. Using $t_{in}(\sigma) = -\sigma/\tilde{u}(0) = -\sigma/U$ with $U=1$ in the dimensionless formulation we have
$\partial_t \tilde{f}(\sigma,t) = \partial_t \hat{f}(t+\sigma) =  \partial_\zeta\hat{f}(\zeta) = \partial_\sigma\hat{f}(t+\sigma) = \partial_\sigma \tilde{f}(\sigma,t)$
and therefore
\begin{align*}
u(\Phi(\sigma,t)) = \partial_t \Phi(\sigma,t) = \partial_\sigma \Phi(\sigma,t) = j(\sigma,t)
\end{align*}
for all $\sigma,t$.
Applying these concepts, regarding the elongation $ \hat{e}(\zeta(\sigma))= \tilde{e}(\sigma,t) = j(\sigma,t)$ and dropping the label $\hat{~}$, the model for the jet's whipping in the Lagrangian description reads
\begin{alignat*}{2}
T^{-1}\mathsf{R}\cdot\partial_\zeta\mathsf{\breve{r}} =& e\mathsf{e_3},\\
T^{-1}\partial_\zeta\mathsf{R} =& -\mathsf{\kappa} \times\mathsf{R},\\
T^{-1}\partial_\zeta\mathsf{\kappa} =& \frac{4}{3} e^3 \mathsf{P}_{3/2}\cdot\mathsf{m} + \Omega\mathsf{\kappa}\times\mathsf{e_3},\\
T^{-1}\partial_\zeta e =& \frac{1}{3} e^2 n_3,\\
T^{-1}\partial_\zeta \mathsf{n} =& -\mathsf{\kappa} \times \mathsf{n} + \mathrm{Re}~e~\big(\mathsf{\kappa} \times \mathsf{e_3} + \frac{1}{3} e n_3 \mathsf{e_3} \big) + 2\mathrm{Re}\Omega e (\mathsf{R}\cdot\mathsf{e_3})\times\mathsf{e_3}\\
& + \mathrm{Re}\Omega^2\mathsf{R}\cdot(\mathsf{e_3} \times (\mathsf{e_3}  \times \mathsf{\breve{r}})) - \mathsf{f}_{ca} - \mathsf{f}_{el} - \mathsf{f}_{air},\\
T^{-1}\partial_\zeta\mathsf{m} =& -\mathsf{\kappa}\times\mathsf{m} + \frac{4}{\epsilon^2}e\mathsf{n}\times\mathsf{e_3} + \frac{\mathrm{Re}}{3}\bigg(e^2\mathsf{P}_3\cdot\mathsf{m} - \frac{1}{4}n_3\mathsf{P}_2\cdot\mathsf{\kappa} \bigg)\\
&-\frac{\mathrm{Re}}{4}\Omega\mathsf{P}_2\cdot\bigg(\frac{1}{3}\mathsf{R}\cdot\mathsf{e_3} n_3 - \frac{1}{3}\mathsf{e_3} n_3 + \bigg(\frac{\mathsf{\kappa}}{e} - \frac{\Omega}{e}\mathsf{e_3}\bigg)\times\mathsf{R}\cdot\mathsf{e_3}\bigg)\\
&-\frac{\mathrm{Re}}{4}\bigg(\frac{1}{e}\mathsf{P}_2\cdot(\mathsf{\kappa}-\Omega\mathsf{e_3}+\Omega\mathsf{R}
\cdot\mathsf{e_3})\bigg)\times(\mathsf{\kappa}-\Omega\mathsf{e_3}+
\Omega\mathsf{R}\cdot\mathsf{e_3}),
\end{alignat*}
with the outer forces
\begin{align*}
\mathsf{f}_{ca} &= \Gamma \frac{1}{\sqrt{e}}\bigg(2\mathsf{\kappa}\times\mathsf{e_3}-\frac{1}{3}e n_3\mathsf{e_3}\bigg),\\
\mathsf{f}_{el} &= \Xi\bigg(4\mathsf{R}\cdot\mathsf{e_3} - \Theta\frac{1}{e^2}\log\bigg(\frac{2}{\epsilon}\sqrt{e}\mathsf{\kappa}\times\mathsf{e_3}\bigg)\bigg),\\
\mathsf{f}_{air} &= \mathrm{M}\mathrm{Re}~e^{3/2}\mathsf{F}\bigg(e\mathsf{e_3},-\mathrm{Re}_\star \frac{1}{\sqrt{e}}(e\mathsf{e_3} + \Omega\mathsf{R}\cdot(\mathsf{e_3}\times\mathsf{\breve{r}}))\bigg),
\end{align*}
and boundary conditions
\begin{alignat*}{4}
\mathsf{\breve{r}}(0) &= 0,\qquad \mathsf{R}(0) &= \mathsf{P}_1,\qquad \mathsf{\kappa}(0) &= 0,\qquad e(0) &= 1,\\
\breve{r}_1(1) &= 0, \qquad \breve{r}_3(1) &= 1, \qquad \mathsf{n}(1) &= 0,\qquad \mathsf{m}(1) &= 0.
\end{alignat*}
The dimensionless end time is $T= t_{end}U/H$ with $t_{end}$ addressing the unknown time of the material point reaching the collector at height $H$.

\begin{figure}[t]
\includegraphics[width=0.49\textwidth]{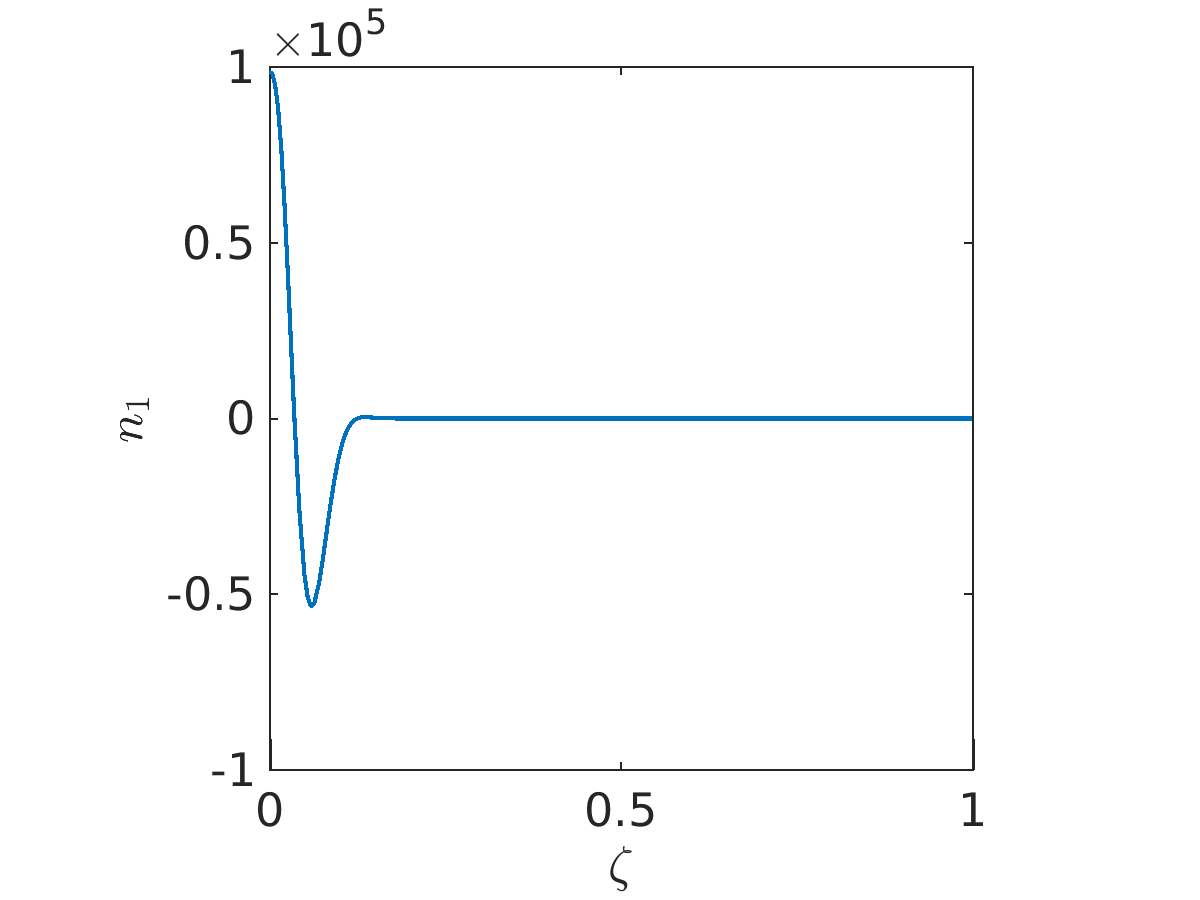}
\includegraphics[width=0.49\textwidth]{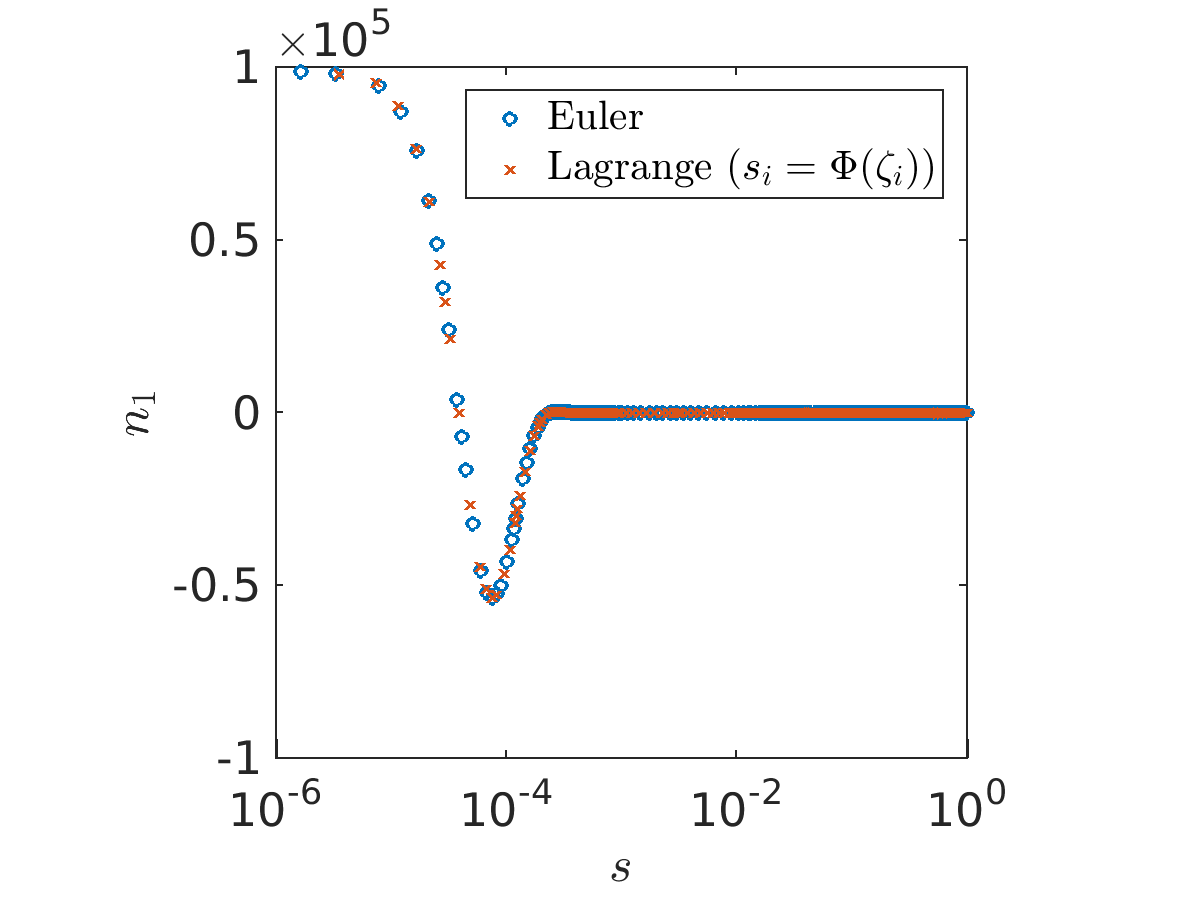}
\caption{\label{fig:compareLayer} Boundary layer in the contact force component $n_1$ for Example~2 with $\epsilon = 0.17$. \emph{Left:} Zoom into the nozzle region in the Lagrangian setting. \emph{Right:} Distribution of discretization points in the Eulerian and Lagrangian setting.}
\end{figure}

The Lagrangian description allows to zoom into the region at the nozzle, where the boundary layers are present for distinct whipping set-ups (see Fig. \ref{fig:compareLayer} (left) for the parameters of Example~2 and $\epsilon = 0.17$). Mapping the Lagrangian parameters $\zeta_i$ used in the simulation as grid points onto the associated Eulerian arc length parameters $s_i = \hat{\Phi}(\zeta_i)$ by means of the nonlinear transformation $\hat{\Phi}$, we see that both descriptions resolve the boundary layer in a similar good manner, Fig.~\ref{fig:compareLayer} (right). The reason why we generally prefer to use the Eulerian description is the desired consideration of high jet thinning which goes along with the development of boundary layers at the jet end in the Lagrangian description. The transformation from spatial to material description shifts layers from the nozzle region to the jet end.
 
\subsection*{Acknowledgments}
J.\ Rivero-Rodr\'iguez and M.\ P\'erez-Saborid would like to thank Profs.\ A.\ Fern\'andez-Nieves, and J.\ Guerrero-Mill\'an for many helpful discussions. This work has been supported by the Ministry of Science and Innovation of Spain (project DPI 2010-20450-C03-02) and the German Research Foundation (DFG, project 251706852, MA 4526/2-1, WE 2003/4-1).

\end{document}